\documentclass[english]{article}
\usepackage[T1]{fontenc}
\usepackage[latin9]{inputenc}
\usepackage{geometry}
\usepackage{units}
\usepackage{textcomp}
\usepackage{amsmath}
\usepackage{amssymb}
\usepackage{graphicx}

\makeatletter

\providecommand{\tabularnewline}{\\}

\@ifundefined{date}{}{\date{}}
\usepackage{nicefrac}

\usepackage{subcaption}
\usepackage{tikz}
\usepackage{pgf}
\usepackage{pgfplots}
\usepackage{placeins}

\usetikzlibrary{arrows}
\usepackage{cite}

\usepackage{babel}

\makeatother

\usepackage{babel}
\begin{document}
\title{Contextuality Analysis of Impossible Figures}
\author{Víctor H. Cervantes \& Ehtibar N. Dzhafarov\\
\\
Purdue University}
\maketitle
\begin{abstract}
This paper has two purposes. One is to demonstrate contextuality analysis
of systems of epistemic random variables. The other is to evaluate
the performance of a new, hierarchical version of the measure of (non)contextuality
introduced in earlier publications. As objects of analysis we use
impossible figures of the kind created by the Penroses and Escher.
We make no assumptions as to how an impossible figure is perceived,
taking it instead as a fixed physical object allowing one of several
deterministic descriptions. Systems of epistemic random variables
are obtained by probabilistically mixing these deterministic systems.
This probabilistic mixture reflects our uncertainty or lack of knowledge
rather than random variability in the frequentist sense. 
\end{abstract}

\section{Introduction}

Our main purpose is to illustrate the use of epistemic random variables
using objects that are naturally described in a deterministic way,
but not uniquely. That is, these objects are described by one of several
deterministic systems. The first applications of contextuality analysis
to such systems is presented in \cite{Dzh_CbDSTC}, using various
deterministic representations of the Liar's paradox. In the present
paper the objects of the analysis are the so-called impossible figures:
the Penrose triangle and several similar figures, as well as the Ascending
and Descending staircase lithograph by M.\ C.\ Escher. The triangle
and the staircase figures were famously discussed by Penrose and Penrose
\cite{Penrose}. In Appendix B to this paper we report several measures
of contextuality computed for our systems of epistemic random variables,
but in the main text we focus on one measure only, introduced here
for the first time, a hierarchical version of the (non)contextuality
measure $\text{CNT}_{2}$-$\text{NCNT}_{2}$ described in \cite{KujDzh_Measures,DzhKujCer_Polytope}.

The Contextuality-by-Default (CbD) theory \cite{Dzh_CbDSTC,KujDzh_Measures,DzhKujCer_Polytope,DzhCerKuj_Canonical}
has been developed to apply to abstract systems of random variables,
irrespective of one's interpretation of probabilistic notions involved.
However, most of its applications have dealt with empirical data and
random variables understood in the frequentist sense. In this paper
we use the term \emph{epistemic random variable} to denote a variable
for which the probabilities with which it falls in different sets
of possible values reflect our uncertainty or lack of knowledge rather
than random variability in the frequentist sense.

A system $\mathcal{R}$ of random variables is a set of double-indexed
random variables $R_{q}^{c}$, where $q\in Q$ denotes their \emph{content},
that can be defined as a question to which the random variable responds,
and $c\in C$ is their \emph{context}, encompassing the conditions
under which it is recorded. A system can be presented as

\begin{equation}
\mathcal{R}=\{R_{q}^{c}:c\in C,q\in Q,q\prec c\},
\end{equation}
where $q\prec c$ indicates that content $q$ is responded to in context
$c$. The variables of the subset

\begin{equation}
R^{c}=\{R_{q}^{c}:q\in Q,q\prec c\}
\end{equation}
are \emph{jointly distributed}, whereas any two random variables in
the subset

\begin{equation}
\mathcal{R}_{q}=\{R_{q}^{c}:c\in C,q\prec c\}
\end{equation}
are \emph{stochastically unrelated.}\footnote{More generally, any two $R_{q}^{c},R_{q'}^{c'}\in\mathcal{R}$ with
$c\neq c'$ are stochastically unrelated. That is, they do not possess
a joint distribution.} The set $R^{c}$ is called the \emph{bunch} corresponding to context
$c$ and the set $\mathcal{R}_{q}$ is referred to as the \emph{connection}
for content $q$.

We will limit our discussion and applications to finite systems of
binary random variables, with $n=\left|Q\right|$ and $m=\left|C\right|$.
Without loss of generality we may assume all variables $R_{q}^{c}$
take values $0/1$. These systems can be represented by three vectors.
The first one is 
\begin{eqnarray}
\mathbf{l} & = & (1,p_{q}^{c}:q\in Q,c\in C,q\prec c)^{\intercal},
\end{eqnarray}
the vector of the low-level marginals, $p_{q}^{c}=\langle R_{q}^{c}\rangle=\Pr(R_{q}^{c}=1)$,
with 1 formally equal to $\langle\cdot\rangle$. The second vector
is 
\begin{eqnarray}
\mathbf{c} & = & (\min(p_{q}^{c_{1}},p_{q}^{c_{2}}):q\in Q,c_{1},c_{2}\in C,q\prec c_{1},c_{2},c_{1}\neq c_{2}),
\end{eqnarray}
the vector of maximal probabilities with which two random variables
in a connection could both equal $1$ (if they possessed a joint distribution).
It is uniquely determined by vector $\mathbf{l}$. The third vector
is

\begin{equation}
\mathbf{b}=(\mathbf{b}_{2},\ldots,\mathbf{b}_{r})^{\intercal},
\end{equation}
where

\begin{equation}
\mathbf{b}_{s}=(p_{q_{1},\ldots,q_{s}}^{c}:q_{1},\ldots,q_{s}\in Q,c\in C,q_{1},\ldots,q_{s}\prec c,q_{1},\ldots,q_{s}\text{ are distinct}).
\end{equation}
Here, $s=2,\ldots,r$, with $p_{q_{1},\ldots,q_{s}}^{c}=\langle R_{q_{1}}^{c}\cdots R_{q_{s}}^{c}\rangle=\Pr(R_{q_{1}}^{c}=1,\ldots,R_{q_{s}}^{c}=1)$,
and $2\leq r\leq n$ is the largest number of distinct $q$'s within
a bunch.\footnote{We could allow $r=1$, in which case $\mathbf{b}$ is empty and the
system is trivially noncontextual.} Thus, $\mathbf{b}$ is a minimal set of probabilities that completely
describes the joint distributions of the bunches in the system (for
a given vector $\mathbf{l}$). The coordinates of a specific system
of random variables are then given by

\begin{equation}
\mathbf{p}^{*}=(\mathbf{l}^{*},\mathbf{c}^{*},\mathbf{b}^{*})^{\intercal}.
\end{equation}
This is the \emph{reduced vectorial representation} of the system,
as discussed in \cite{KujDzh_Measures}. The system is noncontextual
if and only if there is a vector $\mathbf{h}\geq0$ (component-wise)
such that

\begin{equation}
\mathbf{Mh}=\mathbf{p}^{*}.\label{eq:LFT}
\end{equation}
{ Here, the elements of $\mathbf{h}$ are probabilities assigned
to all possible combinations of values ($1$'s and $0$'s) assigned
to all random variables $R_{q}^{c}$ in the system, and $\mathbf{M}$
is an incidence (Boolean) matrix \cite{KujDzh_Measures,DzhCerKuj_Canonical}:
in the row of $\mathbf{M}$ corresponding to a given element of $\mathbf{p}^{*}$,
say, $\Pr(R_{q_{1}}^{c}=\ldots=R_{q_{s}}^{c}=1)$, we put 1 in the
columns corresponding to the elements of $\mathbf{h}$ in which $R_{q_{1}}^{c},\ldots,R_{q_{s}}^{c}$
are assigned the value $1$; other columns in this row are filled
with $0$. } Denoting the rows of $\mathbf{M}$ that correspond to
$\mathbf{l}^{*},\mathbf{c}^{*},\mathbf{b}^{*}$ by, respectively,
$\mathbf{M_{l}},\mathbf{M_{c}},\mathbf{M_{b}}$, we can rewrite (\ref{eq:LFT})
in extenso:

\begin{equation}
\left(\mathbf{M}_{\mathbf{l}},\mathbf{M}_{\mathbf{c}},\mathbf{M}_{\mathbf{b}}\right)^{\intercal}\mathbf{h}=(\mathbf{l}^{*},\mathbf{c}^{*},\mathbf{b}^{*})^{\text{\ensuremath{\intercal}}}.\label{eq:extensoLFT}
\end{equation}

Ref. \cite{KujDzh_Measures} also introduces a measure of contextuality,
$\text{CNT}_{2}$, with its natural extension into a measure of noncontextuality,
$\text{NCNT}_{2}$. In \cite{DzhKujCer_Polytope}, the behavior of
both measures was characterized for a special class of systems, known
as cyclic. These measures can be computed with the aid of linear programming.
To compute $\text{CNT}_{2}$, one solves the following task 
\begin{center}
\begin{tabular}{|ccc|}
\hline 
{\footnotesize{}{}find}  & {\footnotesize{}{}minimizing}  & {\footnotesize{}{}subject to}\tabularnewline
\hline 
{\small{}{}$\mathbf{x}$}  & {\small{}{}$\mathbf{1}^{\intercal}\mathbf{d}$}  & {\small{}{}$-\mathbf{d}\leq\mathbf{b}^{*}-\mathbf{M_{b}x}\leq\mathbf{d}$}\tabularnewline
 &  & {\small{}{}$\mathbf{x},\mathbf{d}\geq0$}\tabularnewline
 &  & {\small{}{}$\mathbf{M_{l}x}=\mathbf{l}^{*}$}\tabularnewline
 &  & {\small{}{}$\mathbf{M_{c}x}=\mathbf{c}^{*}$}\tabularnewline
\hline 
\end{tabular}, 
\par\end{center}

and, for any solution $\mathbf{x}^{*}$, one finds $\text{CNT}_{2}=\left\Vert \mathbf{b^{*}-\mathbf{M_{b}x^{*}}}\right\Vert _{1}$
($L_{1}$-norm). By enumerating the elements of $\mathbf{b}^{*}$
as $1,\ldots,K$, the $\text{NCNT}_{2}$ measure of noncontextuality
is computed as

\begin{equation}
\text{NCNT}_{2}=\min_{i=1,\ldots,K}\{\min(d_{i}^{*-},d_{i}^{*+})\},
\end{equation}
where $d_{i}^{*-},d_{i}^{*+}$ are solutions of the following linear
programming tasks (denoting by $\mathbf{e}_{i}$ the unit vector with
unity in its $i$-th element) 
\begin{center}
\begin{tabular}{|ccc|}
\hline 
{\footnotesize{}{}find}  & {\footnotesize{}{}maximizing}  & {\footnotesize{}{}subject to}\tabularnewline
\hline 
{\small{}{}$\mathbf{x},d_{i}^{-}$}  & {\small{}{}$d_{i}^{-}$}  & {\small{}{}$\mathbf{b}^{*}-d_{i}^{-}\mathbf{e}_{i}=\mathbf{M_{b}x}$}\tabularnewline
 &  & {\small{}{}$\mathbf{x},d_{i}^{-}\geq0$}\tabularnewline
 &  & {\small{}{}$\mathbf{M_{l}x}=\mathbf{l}^{*}$}\tabularnewline
 &  & {\small{}{}$\mathbf{M_{c}x}=\mathbf{c}^{*}$}\tabularnewline
\hline 
\end{tabular}%
\begin{tabular}{|ccc|}
\hline 
{\footnotesize{}{}find}  & {\footnotesize{}{}maximizing}  & {\footnotesize{}{}subject to}\tabularnewline
\hline 
{\small{}{}$\mathbf{x},d_{i}^{+}$}  & {\small{}{}$d_{i}^{+}$}  & {\small{}{}$\mathbf{b}^{*}+d_{i}^{+}\mathbf{e}_{i}=\mathbf{M_{b}x}$}\tabularnewline
 &  & {\small{}{}$\mathbf{x},d_{i}^{+}\geq0$}\tabularnewline
 &  & {\small{}{}$\mathbf{M_{l}x}=\mathbf{l}^{*}$}\tabularnewline
 &  & {\small{}{}$\mathbf{M_{c}x}=\mathbf{c}^{*}$}\tabularnewline
\hline 
\end{tabular}$\:$. 
\par\end{center}

Clearly, for a contextual system $\text{CNT}_{2}>0$ and $\text{NCNT}_{2}=0$,
whereas for a noncontextual one, $\text{CNT}_{2}=0$ and $\text{NCNT}_{2}\geq0$.

\section{Hierarchically measuring contextuality}

The development of this measure has greatly benefited from discussions
with and critical analysis by Janne V. Kujala (see Acknowledgements).

For a given vector $\mathbf{l}^{*}$ (hence also the vector $\mathbf{c}^{*}$),
we call the convex polytope

\begin{equation}
\mathbb{K}=\{\mathbf{b}|\exists\mathbf{h}\geq0:\mathbf{\left(\mathbf{M}_{\mathbf{l}},\mathbf{M}_{\mathbf{c}},\mathbf{M}_{\mathbf{b}}\right)^{\intercal}h}=(\mathbf{l^{*},\mathbf{c^{*},\mathbf{b}}})^{\intercal}\}
\end{equation}
the \emph{noncontextuality} polytope. A system represented by the
point $\mathbf{p}^{*}=(\mathbf{l}^{*},\mathbf{c}^{*},\mathbf{b}^{*})^{\intercal}$
is noncontextual if and only if $\mathbf{b}^{*}\in\mathbb{K}$. The
measures $\text{CNT}_{2}$ and $\text{NCNT}_{2}$ are the $L_{1}$-distance
from $\mathbf{b}^{*}$ to the surface of $\mathbb{K}$ when the system
is, respectively, contextual or noncontextual.

In \cite{DzhKujCer_Polytope,KujDzh_Proof} it is noted that

\begin{equation}
\max(0,p_{q_{1}}^{c}+p_{q_{2}}^{c}-1)\leq p_{q_{1},q_{2}}^{c}\leq\min(p_{q_{1}}^{c},p_{q_{2}}^{c}).
\end{equation}
This observation is easily generalized as

\begin{equation}
\max\left(0,\frac{p_{q_{1},q_{2}}^{c}+p_{q_{1},q_{3}}^{c}+p_{q_{2},q_{3}}^{c}-1}{2}\right)\leq p_{q_{1},q_{2},q_{3}}^{c}\leq\min(p_{q_{1},q_{2}}^{c},p_{q_{1},q_{3}}^{c},p_{q_{2},q_{3}}^{c}),
\end{equation}
\begin{equation}
\begin{array}{r}
\max\left(0,\frac{p_{q_{1},q_{2},q_{3}}^{c}+p_{q_{1},q_{2},q_{4}}^{c}+p_{q_{1},q_{3},q_{4}}^{c}+p_{q_{2},q_{3},q_{4}}^{c}-1}{3}\right)\leq p_{q_{1},q_{2},q_{3},q_{4}}^{c}\\
\leq\min(p_{q_{1},q_{2},q_{3}}^{c},p_{q_{1},q_{2},q_{4}}^{c},p_{q_{1},q_{3},q_{4}}^{c},p_{q_{2},q_{3},q_{4}}^{c}),
\end{array}
\end{equation}
and in complete generality,

\begin{equation}
\begin{array}{r}
\max\left(0,\frac{1}{s-1}\left(\sum_{k=1}^{s}p_{\{q_{1},\ldots,q_{s}\}\setminus\{q_{k}\}}^{c}-1\right)\right)\leq p_{q_{1},\ldots,q_{s}}^{c}\\
\leq\min(p_{\{q_{1},\ldots,q_{s}\}\setminus\{q_{k}\}}^{c}:k=1,\ldots,s),
\end{array}
\end{equation}
for $s=2,\ldots,r$. That is, the elements of $\mathbf{b}^{*}$ are
hierarchically bounded, with $\mathbf{l}^{*}$ providing the bounds
for $\mathbf{b}_{2}^{*}$ and $\mathbf{b}_{s-1}^{*}$ determining
the bounds for $\mathbf{b}_{s}^{*}$ if $2<s\leq r$.

These hierarchical restrictions suggest a hierarchical way of approaching
the measurement of the (non)contextuality of a system $\mathcal{R}$
represented by $\mathbf{p}^{*}$. Consider the systems of equations
\begin{eqnarray}
\left(\mathbf{M}_{\mathbf{l}},\mathbf{M}_{\mathbf{c}},\mathbf{M}_{2}\right)^{\intercal}\mathbf{h}_{2} & = & \left(\mathbf{l}^{*},\mathbf{c}^{*},\mathbf{b}_{2}^{*}\right)^{\intercal}\nonumber \\
 & \vdots\nonumber \\
\left(\mathbf{M}_{\mathbf{l}},\mathbf{M}_{\mathbf{c}},\mathbf{M}_{2},\ldots,\mathbf{M}_{s}\right)^{\intercal}\mathbf{h}_{s} & = & \left(\mathbf{l}^{*},\mathbf{c}^{*},\mathbf{b}_{2}^{*},\ldots,\mathbf{b}_{s}^{*}\right)^{\intercal}\label{eq:systems}\\
 & \vdots\nonumber \\
\left(\mathbf{M}_{\mathbf{l}},\mathbf{M}_{\mathbf{c}},\mathbf{M}_{2},\ldots,\mathbf{M}_{r}\right)^{\text{\ensuremath{\intercal}}}\mathbf{h}_{r} & = & \left(\mathbf{l}^{*},\mathbf{c}^{*},\mathbf{b}_{2}^{*},\ldots,\mathbf{b}_{r}^{*}\right)^{\text{\ensuremath{\intercal}}}=\left(\mathbf{l}^{*},\mathbf{c}^{*},\mathbf{b}^{*}\right)^{\intercal}=\mathbf{M}\mathbf{h},\nonumber 
\end{eqnarray}
where $\mathbf{M}_{s}$ is the submatrix formed by the rows of $\mathbf{M}$
corresponding to the elements of $\mathbf{b}_{s}^{*}$. Clearly, for
a contextual system there is a value $2\leq s^{*}\leq r$ such that
there is no solution $\mathbf{h}_{s}\geq0$ for any $s\geq s^{*}$
while there is a solution for each $s<s^{*}$. For a noncontextual
system, the solution to the system (\ref{eq:LFT}) implies the solution
to all systems (\ref{eq:systems}). Therefore, if $\mathcal{R}$ is
contextual, we can further qualify its contextuality and say that
it is contextual at level $s^{*}$. Its degree of contextuality at
level $s^{*}$ can be computed by solving the linear programming task 
\begin{center}
\begin{tabular}{|ccc|}
\hline 
{\footnotesize{}{}find}  & {\footnotesize{}{}minimizing}  & {\footnotesize{}{}subject to}\tabularnewline
\hline 
{\small{}{}$\mathbf{x}$}  & {\small{}{}$\mathbf{1}^{\intercal}\mathbf{d}$}  & {\small{}{}$-\mathbf{d}\leq\mathbf{b}_{s^{*}}^{*}-\mathbf{M}_{s^{*}}\mathbf{x}\leq\mathbf{d}$}\tabularnewline
 &  & {\small{}{}$\mathbf{x},\mathbf{d}\geq0$}\tabularnewline
 &  & {\small{}{}$\mathbf{M_{l}x}=\mathbf{l}^{*}$}\tabularnewline
 &  & {\small{}{}$\mathbf{M_{c}x}=\mathbf{c}^{*}$}\tabularnewline
 &  & {\small{}{}$\mathbf{M}_{2}\mathbf{x}=\mathbf{b}_{2}^{*}$}\tabularnewline
 &  & {\small{}{}$\vdots$}\tabularnewline
 &  & {\small{}{}$\mathbf{M}_{s^{*}-1}\mathbf{x}=\mathbf{b}_{s^{*}-1}^{*}$}\tabularnewline
\hline 
\end{tabular}$\:$, 
\par\end{center}

and computing, for any solution $\mathbf{x^{*}}$,

\begin{equation}
\text{CNT}_{2}^{s^{*}}=\left\Vert \mathbf{b}_{s^{*}}^{*}-\mathbf{M}_{s^{*}}\mathbf{x^{*}}\right\Vert _{1}.
\end{equation}
Moreover, for each $s$ for which $\left(\mathbf{M}_{\mathbf{l}},\mathbf{M}_{\mathbf{c}},\mathbf{M}_{2},\ldots,\mathbf{M}_{s}\right)^{\intercal}\mathbf{h}_{s}=\left(\mathbf{l}^{*},\mathbf{c}^{*},\mathbf{b}_{2}^{*},\ldots,\mathbf{b}_{s}^{*}\right)^{\intercal}$
has a solution, we can compute the noncontextuality of $\mathcal{R}$
at level $s$ as

\begin{equation}
\text{NCNT}_{2}^{s}=\min_{i=1,\ldots,K_{s}}\{\min(d_{i}^{*-},d_{i}^{*+})\},
\end{equation}
where $K_{s}$ is the number of elements of $\mathbf{b}_{s}^{*}$,
and $d_{i}^{*-},d_{i}^{*+}$ are solutions to the linear programming
tasks 
\begin{center}
\begin{tabular}{|ccc|}
\hline 
{\footnotesize{}{}find}  & {\footnotesize{}{}maximizing}  & {\footnotesize{}{}subject to}\tabularnewline
\hline 
{\small{}{}$\mathbf{x},d_{i}^{-}$}  & {\small{}{}$d_{i}^{-}$}  & {\small{}{}$\mathbf{b}_{s}^{*}-d_{i}^{+}\mathbf{e}_{i}=\mathbf{M}_{s}\mathbf{x}$}\tabularnewline
 &  & {\small{}{}$\mathbf{x},d_{i}^{-}\geq0$}\tabularnewline
 &  & {\small{}{}$\mathbf{M_{l}x}=\mathbf{l}^{*}$}\tabularnewline
 &  & {\small{}{}$\mathbf{M_{c}x}=\mathbf{c}^{*}$}\tabularnewline
 &  & {\small{}{}$\mathbf{M}_{2}\mathbf{x}=\mathbf{b}_{2}^{*}$}\tabularnewline
 &  & {\small{}{}$\vdots$}\tabularnewline
 &  & {\small{}{}$\mathbf{M}_{s-1}\mathbf{x}=\mathbf{b}_{s-1}^{*}$}\tabularnewline
\hline 
\end{tabular}%
\begin{tabular}{|ccc|}
\hline 
{\footnotesize{}{}find}  & {\footnotesize{}{}maximizing}  & {\footnotesize{}{}subject to}\tabularnewline
\hline 
{\small{}{}$\mathbf{x},d_{i}^{+}$}  & {\small{}{}$d_{i}^{+}$}  & {\small{}{}$\mathbf{b}_{s}^{*}+d_{i}^{+}\mathbf{e}_{i}=\mathbf{M}_{s}\mathbf{x}$}\tabularnewline
 &  & {\small{}{}$\mathbf{x},d_{i}^{+}\geq0$}\tabularnewline
 &  & {\small{}{}$\mathbf{M_{l}x}=\mathbf{l}^{*}$}\tabularnewline
 &  & {\small{}{}$\mathbf{M_{c}x}=\mathbf{c}^{*}$}\tabularnewline
 &  & {\small{}{}$\mathbf{M}_{2}\mathbf{x}=\mathbf{b}_{2}^{*}$}\tabularnewline
 &  & {\small{}{}$\vdots$}\tabularnewline
 &  & {\small{}{}$\mathbf{M}_{s-1}\mathbf{x}=\mathbf{b}_{s-1}^{*}$}\tabularnewline
\hline 
\end{tabular}$\:$. 
\par\end{center}

In this way, we construct the hierarchical measure of (non)contextuality
which characterizes the degree of (non)contextuality of a system $\mathcal{R}$
by a vector of size $s^{*}-1$ if the system is contextual, or of
size $r-1$ if the system is noncontextual. For a contextual system,
$\text{CNT}_{2}^{s^{*}}$ gives the $L_{1}$-distance from $\mathbf{b}_{s^{*}}^{*}$
to the surface of the polytope

\begin{equation}
\mathbb{K}_{s^{*}}=\begin{cases}
\{\mathbf{b}_{2}|\exists\mathbf{h}\geq0:\left(\mathbf{M}_{\mathbf{l}},\mathbf{M}_{\mathbf{c}},\mathbf{M}_{2}\right)^{\intercal}\mathbf{h}_{2}=\\
\qquad\qquad\qquad\left(\mathbf{l}^{*},\mathbf{c}^{*},\mathbf{b}_{2}\right)^{\intercal}\} & \text{if }s^{*}=2\\
\\
\{\mathbf{b}_{s^{*}}|\exists\mathbf{h}\geq0:\left(\mathbf{M}_{\mathbf{l}},\mathbf{M}_{\mathbf{c}},\mathbf{M}_{2},\ldots,\mathbf{M}_{s^{*}}\right)^{\intercal}\mathbf{h}_{s^{*}}=\\
\qquad\qquad\qquad\left(\mathbf{l}^{*},\mathbf{c}^{*},\mathbf{b}_{2}^{*},\ldots,\mathbf{b}_{s^{*}-1}^{*},\mathbf{b}_{s^{*}}\right)^{\intercal}\} & \text{if }2<s^{*}\leq r
\end{cases}.
\end{equation}
And, at each level $2\leq s<s^{*}$ ($2\leq s\leq r$ for a noncontextual
system), $\text{NCNT}_{2}^{s}$ is the $L_{1}$-distance from $\mathbf{b}_{s}^{*}$
to the surface of the polytope

\begin{equation}
\mathbb{K}_{s}=\{\mathbf{b}_{s}|\exists\mathbf{h}\geq0:\left(\mathbf{M}_{\mathbf{l}},\mathbf{M}_{\mathbf{c}},\mathbf{M}_{2},\ldots,\mathbf{M}_{s}\right)^{\intercal}\mathbf{h}_{s}=\left(\mathbf{l}^{*},\mathbf{c}^{*},\mathbf{b}_{2}^{*},\ldots,\mathbf{b}_{s-1}^{*},\mathbf{b}_{s}\right)^{\intercal}\}.
\end{equation}
Analogously to $\text{CNT}_{2}$ and $\text{NCNT}_{2}$, we have that
for a system contextual at level $s^{*}$, $\text{CNT}_{2}^{s^{*}}>0$
and $\text{NCNT}_{2}^{s^{*}}=0$, and in addition for $s<s^{*}$,
$\text{CNT}_{2}^{s}=0$.

\section{The Penrose triangle}

We will now apply the measures just constructed to {some} drawings
known as impossible figures. The general idea {underlying their contextuality
analysis} is that the epistemic random variables representing an
impossible figure should always be compared to those representing
a realizable (i.e., ``normal'') figure, and the degree of contextuality
be derived from the difference between the two.

\begin{figure}[bt]
\begin{subfigure}[t]{0.5\textwidth} 
\begin{centering}
\caption{\raggedright\label{fig:Penrose-tribar.-1}}
\begin{tikzpicture}[scale = 0.55, line cap = round, line join = round, >= triangle 45, x = 1cm, y = 1cm]


\fill[fill = black, fill opacity = 0.1] (5.5, 6.06) -- (6.5, 6.06) -- (4, 1.73) -- (7, 1.73) -- (7.5, 0.87) -- (2.5, 0.87) -- cycle;

\fill[fill = black, fill opacity = 0.2] (9.5, 0.87) -- (9, 0) -- (6.5, 4.33) -- (5, 1.73) -- (4, 1.73) -- (6.5, 6.06) -- cycle;

\fill[fill = black, fill opacity = 0.3] (3, 0) -- (2.5, 0.87) -- (7.5, 0.87) -- (6, 3.46) -- (6.5, 4.33) -- (9, 0) -- cycle;


\draw [line width = 0.6pt] (2.5, 0.87) -- (5.5, 6.06);
\draw [line width = 0.6pt] (4,   1.73) -- (6.5, 6.06);
\draw [line width = 0.6pt] (5.5, 6.06) -- (6.5, 6.06);
\draw [line width = 0.6pt] (6.5, 4.33) -- (5,   1.73);

\draw [line width = 0.6pt] (6.5, 4.33) -- (9,   0);
\draw [line width = 0.6pt] (6.5, 6.06) -- (9.5, 0.87);
\draw [line width = 0.6pt] (7.5, 0.87) -- (6,   3.46);
\draw [line width = 0.6pt] (9.5, 0.87) -- (9,   0);

\draw [line width = 0.6pt] (2.5, 0.87) -- (7.5, 0.87);
\draw [line width = 0.6pt] (3,   0)    -- (2.5, 0.87);
\draw [line width = 0.6pt] (4,   1.73) -- (7,   1.73);
\draw [line width = 0.6pt] (9,   0)    -- (3,   0);

\end{tikzpicture} 
\par\end{centering}
\end{subfigure} 
\begin{subfigure}[t]{0.5\textwidth} 
\begin{centering}
\caption{\raggedright\label{fig:Penrose-tribar-with}}
\begin{tikzpicture}[scale = 0.55, line cap=round,line join=round,>=triangle 45,x=1cm,y=1cm]


\fill[fill = black, fill opacity = 0.1] (5.5, 6.06) -- (6.5, 6.06) -- (4, 1.73) -- (7, 1.73) -- (7.5, 0.87) -- (2.5, 0.87) -- cycle;

\fill[fill = black, fill opacity = 0.2] (9.5, 0.87) -- (9, 0) -- (6.5, 4.33) -- (5, 1.73) -- (4, 1.73) -- (6.5, 6.06) -- cycle;

\fill[fill = black, fill opacity = 0.3] (3, 0) -- (2.5, 0.87) -- (7.5, 0.87) -- (6, 3.46) -- (6.5, 4.33) -- (9, 0) -- cycle;


\draw [line width = 0.6pt] (2.5, 0.87) -- (5.5, 6.06);
\draw [line width = 0.6pt] (4,   1.73) -- (6.5, 6.06);
\draw [line width = 0.6pt] (5.5, 6.06) -- (6.5, 6.06);
\draw [line width = 0.6pt] (6.5, 4.33) -- (5,   1.73);

\draw [line width = 0.6pt] (6.5, 4.33) -- (9,   0);
\draw [line width = 0.6pt] (6.5, 6.06) -- (9.5, 0.87);
\draw [line width = 0.6pt] (7.5, 0.87) -- (6,   3.46);
\draw [line width = 0.6pt] (9.5, 0.87) -- (9,   0);

\draw [line width = 0.6pt] (2.5, 0.87) -- (7.5, 0.87);
\draw [line width = 0.6pt] (3,   0)    -- (2.5, 0.87);
\draw [line width = 0.6pt] (4,   1.73) -- (7,   1.73);
\draw [line width = 0.6pt] (9,   0)    -- (3,   0);


\draw [line width = 0.6pt, color = blue]         (4, 1.73) -- (3,   1.73);
\draw (2.7, 1.73) node {\small{0}};
\draw [line width = 0.6pt, color = blue]         (5, 5.20) -- (6,   5.20);
\draw (4.7, 5.20) node {\small{1}};
\draw [line width = 0.6pt, color = blue]         (6, 5.20) -- (6.5, 4.33);
\draw [line width = 0.6pt, dotted, color = blue] (4, 1.73) -- (4.5, 0.87);

\draw [line width = 0.6pt, color = yellow]         (6.5, 4.33) -- (7,   5.20);
\draw (7.3, 5.22) node {\small{0}};
\draw [line width = 0.6pt, color = yellow]         (7.5, 0.87) -- (8.5, 0.87);
\draw [line width = 0.6pt, color = yellow]         (8.5, 0.87) -- (9,   1.73);
\draw (9.3, 1.75) node {\small{1}};
\draw [line width = 0.6pt, dotted, color = yellow] (5.5, 4.33) -- (6.5, 4.33);

\draw [line width = 0.6pt, color = red]         (3.5, 0.87) -- (4,   0);
\draw (4, -0.35) node {\small{1}};
\draw [line width = 0.6pt, color = red]         (3.5, 0.87) -- (4,   1.73);
\draw [line width = 0.6pt, color = red]         (8,   0)    -- (7.5, 0.87);
\draw (8, -0.35) node {\small{0}};
\draw [line width = 0.6pt, dotted, color = red] (7.5, 0.87) -- (8,   1.73);

\end{tikzpicture} 
\par\end{centering}
\end{subfigure} 

\vspace{3mm}

\begin{subfigure}[b]{0.5\textwidth} 
\begin{centering}
\caption{\raggedright\label{fig:Look-through-right}}
\begin{tikzpicture}[scale = 0.55, line cap=round,line join=round,>=triangle 45,x=1cm,y=1cm]


\fill[fill = black, fill opacity = 0.1] (5.5, 6.06) -- (6.5, 6.06) -- (4, 1.73) -- (7, 1.73) -- (7.5, 0.87) -- (2.5, 0.87) -- cycle;

\fill[fill = black, fill opacity = 0.2] (9.5, 0.87) -- (9, 0) -- (6.5, 4.33) -- (5, 1.73) -- (4, 1.73) -- (6.5, 6.06) -- cycle;

\fill[fill = black, fill opacity = 0.3] (3, 0) -- (2.5, 0.87) -- (7.5, 0.87) -- (6, 3.46) -- (6.5, 4.33) -- (9, 0) -- cycle;


\draw [line width = 0.6pt] (2.5, 0.87) -- (5.5, 6.06);
\draw [line width = 0.6pt] (4,   1.73) -- (6.5, 6.06);
\draw [line width = 0.6pt] (5.5, 6.06) -- (6.5, 6.06);
\draw [line width = 0.6pt] (6.5, 4.33) -- (5,   1.73);

\draw [line width = 0.6pt] (6.5, 6.06) -- (7, 5.20);
\draw [line width = 0.6pt] (8.5, 0.87) -- (9,   0);
\draw [line width = 0.6pt] (9.5, 0.87) -- (9,   0);
\draw [line width = 0.6pt] (9.5, 0.87) -- (9,   1.73);

\draw [line width = 0.6pt] (2.5, 0.87) -- (3,   0);
\draw [line width = 0.6pt] (2.5, 0.87) -- (7.5, 0.87);
\draw [line width = 0.6pt] (4,   1.73) -- (9,   1.73);
\draw [line width = 0.6pt] (9,   0)    -- (3,   0);


\draw [line width = 0.6pt, color = blue]         (4, 1.73) -- (3,   1.73);
\draw [line width = 0.6pt, color = blue]         (5, 5.20) -- (6,   5.20);
\draw [line width = 0.6pt, color = blue]         (6, 5.20) -- (6.5, 4.33);
\draw [line width = 0.6pt, dotted, color = blue] (4, 1.73) -- (4.5, 0.87);

\draw [line width = 0.6pt, color = yellow]         (6.5, 4.33) -- (7,   5.20);
\draw [line width = 0.6pt, color = yellow]         (7.5, 0.87) -- (8.5, 0.87);
\draw [line width = 0.6pt, color = yellow]         (8.5, 0.87) -- (9,   1.73);
\draw [line width = 0.6pt, dotted, color = yellow] (5.5, 4.33) -- (6.5, 4.33);

\draw [line width = 0.6pt, color = red]         (3.5, 0.87) -- (4,   0);
\draw [line width = 0.6pt, color = red]         (3.5, 0.87) -- (4,   1.73);
\draw [line width = 0.6pt, color = red]         (8,   0)    -- (7.5, 0.87);
\draw [line width = 0.6pt, dotted, color = red] (7.5, 0.87) -- (8,   1.73);

\end{tikzpicture} 
\par\end{centering}
\end{subfigure} 
\begin{subfigure}[b]{0.5\textwidth} 
\begin{centering}
\caption{\raggedright\label{fig:Right-bar-removed.}}
\begin{tikzpicture}[scale = 0.55, line cap=round,line join=round,>=triangle 45,x=1cm,y=1cm]


\fill[fill = black, fill opacity = 0.1] (5.5, 6.06) -- (6.5, 6.06) -- (4, 1.73) -- (9, 1.73) -- (8.5, 0.87) -- (2.5, 0.87) -- cycle;

\fill[fill = black, fill opacity = 0.2] (7, 5.20) -- (6.5, 4.33) -- (5, 1.73) -- (4, 1.73) -- (6.5, 6.06) -- cycle;
\fill[fill = black, fill opacity = 0.2] (9.5, 0.87) -- (9, 0) -- (8.5, 0.87) -- (9, 1.73) -- cycle;

\fill[fill = black, fill opacity = 0.3] (3, 0) -- (2.5, 0.87) -- (7.5, 0.87) -- (8.5, 0.87) -- (9, 0) -- cycle;


\draw [line width = 0.6pt] (2.5, 0.87) -- (5.5, 6.06);
\draw [line width = 0.6pt] (4,   1.73) -- (6.5, 6.06);
\draw [line width = 0.6pt] (5.5, 6.06) -- (6.5, 6.06);
\draw [line width = 0.6pt] (6.5, 4.33) -- (5,   1.73);

\draw [line width = 0.6pt] (6.5, 6.06) -- (7, 5.20);
\draw [line width = 0.6pt] (8.5, 0.87) -- (9,   0);
\draw [line width = 0.6pt] (9.5, 0.87) -- (9,   0);
\draw [line width = 0.6pt] (9.5, 0.87) -- (9,   1.73);

\draw [line width = 0.6pt] (2.5, 0.87) -- (3,   0);
\draw [line width = 0.6pt] (2.5, 0.87) -- (7.5, 0.87);
\draw [line width = 0.6pt] (4,   1.73) -- (9,   1.73);
\draw [line width = 0.6pt] (9,   0)    -- (3,   0);


\draw [line width = 0.6pt, color = blue]         (4, 1.73) -- (3,   1.73);
\draw [line width = 0.6pt, color = blue]         (5, 5.20) -- (6,   5.20);
\draw [line width = 0.6pt, color = blue]         (6, 5.20) -- (6.5, 4.33);
\draw [line width = 0.6pt, dotted, color = blue] (4, 1.73) -- (4.5, 0.87);

\draw [line width = 0.6pt, color = yellow]         (6.5, 4.33) -- (7,   5.20);
\draw [line width = 0.6pt, color = yellow]         (7.5, 0.87) -- (8.5, 0.87);
\draw [line width = 0.6pt, color = yellow]         (8.5, 0.87) -- (9,   1.73);
\draw [line width = 0.6pt, dotted, color = yellow] (5.5, 4.33) -- (6.5, 4.33);

\draw [line width = 0.6pt, color = red]         (3.5, 0.87) -- (4,   0);
\draw [line width = 0.6pt, color = red]         (3.5, 0.87) -- (4,   1.73);
\draw [line width = 0.6pt, color = red]         (8,   0)    -- (7.5, 0.87);
\draw [line width = 0.6pt, dotted, color = red] (7.5, 0.87) -- (8,   1.73);

\end{tikzpicture} 
\par\end{centering}
\end{subfigure} \caption{\label{fig:Penrose-tribar.} (\textbf{a}) The Penrose triangle. (\textbf{b}){\,}The
Penrose triangle with superimposed cuts{, shown with the corresponding
labels (0 or 1).} (\textbf{c}) Look through the right bar of the
triangle. (\textbf{d}) Right bar removed by its cuts. }
\end{figure}
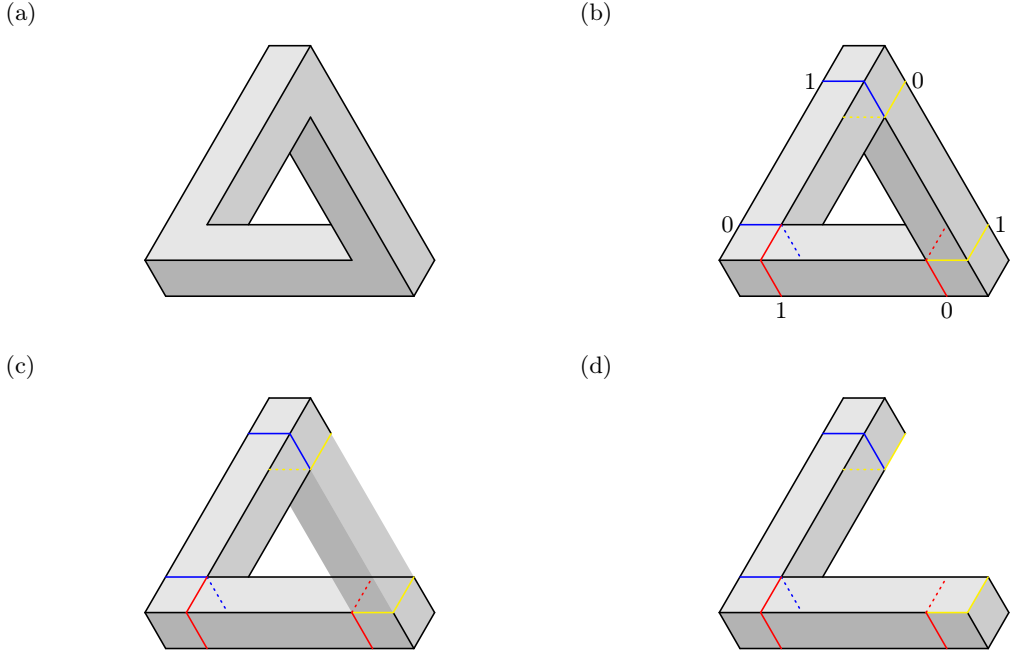

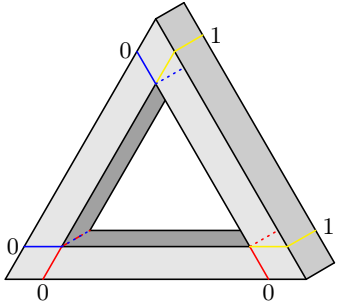
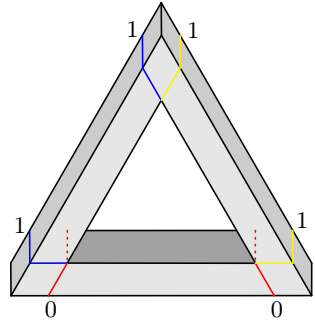
\begin{figure}[tb]
\begin{subfigure}[t]{0.5\textwidth} 
\begin{centering}
\caption{\raggedright\label{fig:Tribar-at-30}}
\begin{tikzpicture}[scale = 0.5, line cap=round,line join=round,>=triangle 45,x=1cm,y=1cm]


\fill[line width=0pt,fill=black,fill opacity=0.1] (-4,0) -- (4,0) -- (0, 6.93) -- cycle;

\fill[line width=0pt,fill=black,fill opacity=0.3] (-2.5, 0.87) -- (2.5, 0.87) -- (0, 5.20) -- cycle;

\fill[line width=0pt,fill=black,fill opacity=0.2] (4.75, 0.44) -- (0.75, 7.36) -- (0,    6.93) -- (4,    0) -- cycle;

\fill[line width=0pt,color=white,fill=white,fill opacity=1] (-1.75, 1.30) -- (2.25,  1.30) -- (0.25,  4.76) -- cycle;



\draw [line width=0.6pt] (-4, 0)    -- (4,0);
\draw [line width=0.6pt] (4,  0)    -- (0,6.93);
\draw [line width=0.6pt] (0,  6.93) -- (-4,0);

\draw [line width=0.6pt] (-2.5, 0.87) -- (2.5, 0.87);
\draw [line width=0.6pt] (2.5,  0.87) -- (0, 5.20);
\draw [line width=0.6pt] (0,    5.20) -- (-2.5, 0.87);


\draw [line width=0.6pt] (4.75, 0.44) -- (0.75, 7.36);
\draw [line width=0.6pt] (0,    6.93) -- (0.75, 7.36);
\draw [line width=0.6pt] (4,    0)    -- (4.75, 0.44);


\draw [line width=0.6pt] (-1.75, 1.30) -- (2.25,  1.30);
\draw [line width=0.6pt] (0.25,  4.76) -- (-1.75, 1.30);


\draw [line width = 0.6pt, color = red]         (-3,   0)    -- (-2.5, 0.87);
\draw (-3, -0.35) node {\small{0}};
\draw [line width = 0.6pt, color = red]         (3,    0)    -- (2.5,  0.87);
\draw (3, -0.35) node {\small{0}};
\draw [line width = 0.6pt, dashed, color = red] (-2.5, 0.87) -- (-1.75, 1.30);
\draw [line width = 0.6pt, dotted, color = red] (2.5,  0.87) -- (3.25, 1.30);

\draw [line width = 0.6pt, color = yellow] (2.5, 0.87) -- (3.5, 0.87);
\draw [line width = 0.6pt, color = yellow] (3.5, 0.87) -- (4.25, 1.30);
\draw (4.6, 1.4) node {\small{1}};
\draw [line width = 0.6pt, color = yellow] (0,   5.20) -- (0.5, 6.06);
\draw [line width = 0.6pt, color = yellow] (0.5, 6.06) -- (1.25, 6.50);
\draw (1.6, 6.5) node {\small{1}};

\draw [line width = 0.6pt, color = blue]         (-3.5,  0.87) -- (-2.5, 0.87);
\draw (-3.8, 0.87) node {\small{0}};
\draw [line width = 0.6pt, color = blue]         (-0.5, 6.06) -- (0, 5.20);
\draw (-0.8, 6.1) node {\small{0}};
\draw [line width = 0.6pt, dotted, color = blue] (-2.5, 0.87) -- (-1.75, 1.30);
\draw [line width = 0.6pt, dotted, color = blue] (0, 5.20) -- (0.75, 5.63);

\end{tikzpicture} 
\par\end{centering}
\end{subfigure} \begin{subfigure}[t]{0.5\textwidth} \caption{\raggedright\label{fig:Tribar-at-90}}

\begin{centering}
\begin{tikzpicture}[scale = 0.5, rotate = -120, line cap=round,line join=round,>=triangle 45,x=1cm,y=1cm]


\fill[line width=0pt,fill=black,fill opacity=0.1] (-3.25, 0.44) -- (4.75, 0.44) -- (0.75, 7.36) -- cycle;

\fill[line width=0pt,fill=black,fill opacity=0.3] (2.25, 1.30) -- (3.25, 1.30) -- (0.75, 5.64) -- (0.25, 4.75) -- cycle;

\fill[line width=0pt,fill=black,fill opacity=0.2] (-4, 0) -- (4, 0) -- (4.75, 0.44) -- (-3.25, 0.44) -- 
(0.75, 7.36) -- (0, 6.93) -- cycle;

\fill[line width=0pt,color=white,fill=white,fill opacity=1] (-1.75, 1.30) -- (2.25, 1.30) -- (0.25, 4.76) -- cycle;



\draw [line width=0.6pt] (4.75, 0.44) -- (0.75, 7.36);
\draw [line width=0.6pt] (-1.75, 1.30) -- (3.25,  1.30);
\draw [line width=0.6pt] (3.25,  1.30) -- (0.75, 5.64);
\draw [line width=0.6pt] (0.75,  5.64) -- (-1.75, 1.30);
\draw [line width=0.6pt] (-3.25, 0.44) -- (4.75, 0.44);
\draw [line width=0.6pt] (0.75, 7.36) -- (-3.25, 0.44);


\draw [line width=0.6pt] (-4, 0) -- (4, 0);
\draw [line width=0.6pt] (0,    6.93) -- (0.75, 7.36);
\draw [line width=0.6pt] (4,    0)    -- (4.75, 0.44);
\draw [line width=0.6pt] (0, 6.93) -- (-4, 0);
\draw [line width=0.6pt] (-4, 0) -- (-3.25, 0.44);


\draw [line width=0.6pt] (2.25, 1.30) -- (0.25, 4.76);


\draw [line width = 0.6pt, color = blue] (-2.25, 0.44) -- (-1.75, 1.30);
\draw [line width = 0.6pt, color = blue] (-3,    0)    -- (-2.25, 0.44);
\draw (-3, -0.3) node {\small{1}};
\draw [line width = 0.6pt, color = blue] (3,     0)    -- (3.75,  0.44);
\draw (3, -0.3) node {\small{1}};
\draw [line width = 0.6pt, color = blue] (3.75,  0.44) -- (3.25, 1.30);

\draw [line width = 0.6pt, color = red]         (0.75, 5.63) -- (1.25, 6.50);
\draw (1.53, 6.75) node {\small{0}};
\draw [line width = 0.6pt, color = red]         (3.25, 1.30) -- (4.25, 1.30);
\draw (4.53, 1.54) node {\small{0}};
\draw [line width = 0.6pt, dotted, color = red] (0,    5.20) -- (0.75, 5.63);
\draw [line width = 0.6pt, dotted, color = red] (2.5,  0.87) -- (3.25, 1.30);

\draw [line width = 0.6pt, color = yellow] (-0.5,  6.06) -- (0.25, 6.50);
\draw (-0.85, 6.15) node {\small{1}};
\draw [line width = 0.6pt, color = yellow] (-2.75, 1.30) -- (-1.75, 1.30);
\draw [line width = 0.6pt, color = yellow] (-3.5,  0.87) -- (-2.75, 1.30);
\draw (-3.8, 1.1) node {\small{1}};
\draw [line width = 0.6pt, color = yellow] (0.25,  6.50) -- (0.75, 5.63);

\end{tikzpicture} 
\par\end{centering}
\end{subfigure} \caption{\label{fig:Realizable-tribars.}Realizable triangles with superimposed
cuts{, shown with the corresponding labels (0 or 1).} (\textbf{a})
Triangle at 30 deg {oblique} projection. (\textbf{b}) Triangle at
90 deg {oblique} projection. }
\end{figure}

We begin with the well-known Penrose triangle, depicted in Figure
\ref{fig:Penrose-tribar.-1}. Observe that one can see precisely two
faces of each of the three bars forming the figure. So each corner
of the triangle is formed by four of these faces, one of which ends
in the inner fold of the corner (invisible in Figure \ref{fig:Penrose-tribar.-1};
in other figures, e.g. in Figure \ref{fig:Tribar-at-30}, if the inner
fold is visible, then two faces, of two different bars, end in it).
Looking at the two faces of a given bar near one of the corners, either
one of the faces ends in the inner fold while another does not (we
encode this case by $0$), or both do not (encoded as $1$). In Figure
\ref{fig:Penrose-tribar-with} this is shown by interrupted and solid
lines of the cuts made in each bar at the two corners it connects.
The endpoint labels $1$ and $0$ correspond to, respectively, the
case when both cut lines are solid, and the case when one of them
is interrupted.

We do not claim that this encoding describes how the figure or its
elements are perceived. The latter is a question for an empirical
investigation, such as the one presented in \cite{Asano}, where contextuality
analysis was applied to perception of an ambiguous figure (SchrÃ¶der's
stair). We are merely selecting a possible description of the figure
as a fixed physical object. { Other mathematical descriptions of
the Penrose triangle and other impossible figures (some of them very
different from those considered in this paper) can be found in \cite{Cowan,Penrose_CoHom,Sugihara}.
} 
Perception may make use of such descriptions, but it is most likely
a complex process with descriptions changing in time.

We can construct a system representation of the Penrose triangle in
two ways. The first one consists in looking at each of the three bars
separately, and arbitrarily picking one of its two ends as the first
one in an ordered pair. We see in Figure \ref{fig:Penrose-tribar.}
that whenever the end we pick is coded 0, the other end of the bar
is coded 1, and vice-versa. Thus, we get two deterministic descriptions,

\begin{equation}
\begin{array}{|c|c||c|}
\hline 0 & 1 & c\\
\hline\hline q & q' & 
\\\hline \end{array}\textnormal{ and \ensuremath{\begin{array}{|c|c||c|}
\hline 1 & 0 & c\\
\hline\hline q & q' & 
\\\hline \end{array}\:},}
\end{equation}
where $c$ designates a specific bar (here, left, right, or bottom
one), and $\left(q,q'\right)$ is an ordered pair of its endpoints.
Since these deterministic descriptions are equally plausible (i.e.,
there is no preferred ordering of the endpoints of an isolated bar),
we can assign the epistemic probability $\nicefrac{1}{2}$ to each
of them, obtaining thereby two perfectly anticorrelated uniformly
distributed epistemic random variables,

\begin{equation}
\begin{array}{c|c|c|c}
 & R_{q'}^{c}=0 & R_{q'}^{c}=1\\
\hline R_{q}^{c}=0 & 0 & \nicefrac{1}{2} & \nicefrac{1}{2}\\
\hline R_{q}^{c}=1 & \nicefrac{1}{2} & 0 & \nicefrac{1}{2}\\
\hline  & \nicefrac{1}{2} & \nicefrac{1}{2}
\end{array}\:.\label{eq:corr=00003D00003D-1}
\end{equation}
The stand-alone numbers in the table are joint and marginal probabilities.

We now have to add for comparison epistemic random variables describing
a bar of a realizable (i.e., ``normal'') triangle, as the ones shown
in Figure \ref{fig:Realizable-tribars.}. {We define a realizable
figure as one that can be viewed as an oblique projection of a physical
object of relatively small thickness (so that perspective can be ignored).}
Each of the contexts representing a realizable bar has two identically
labelled ends, and depending on the {oblique} projection angle,
each bar (left, right, or bottom) can be labeled $\left(1,1\right)$
or $\left(0,0\right)$. Using the same uniform epistemic mixing as
before, we obtain two perfectly correlated uniformly distributed epistemic
random variables,

\begin{equation}
\begin{array}{c|c|c|c}
 & R_{q'}^{c}=0 & R_{q'}^{c}=1\\
\hline R_{q}^{c}=0 & \nicefrac{1}{2} & 0 & \nicefrac{1}{2}\\
\hline R_{q}^{c}=1 & 0 & \nicefrac{1}{2} & \nicefrac{1}{2}\\
\hline  & \nicefrac{1}{2} & \nicefrac{1}{2}
\end{array}\:.\label{eq:corr=00003D00003D1}
\end{equation}
Repeating this reasoning for each of the three bars, we get the context-content
matrix 
\begin{center}
\emph{Option 1} 
\begin{align}
\begin{array}{|c|c|c|c|c|c||c|}
\hline R_{1}^{1} & R_{2}^{1} &  &  &  &  & c_{1}\\
\hline R_{1}^{2} & R_{2}^{2} &  &  &  &  & c_{2}\\
\hline  &  & R_{3}^{3} & R_{4}^{3} &  &  & c_{3}\\
\hline  &  & R_{3}^{4} & R_{4}^{4} &  &  & c_{4}\\
\hline  &  &  &  & R_{5}^{5} & R_{6}^{5} & c_{5}\\
\hline  &  &  &  & R_{5}^{6} & R_{6}^{6} & c_{6}\\
\hline\hline q_{1} & q_{2} & q_{3} & q_{4} & q_{5} & q_{6} & \mathcal{P}_{3}^{1}
\\\hline \end{array}.\label{eq:triangleOption1}
\end{align}
\par\end{center}

The contexts and contents in $\mathcal{P}_{3}^{1}$ have been numbered
so that the odd contexts represent the bars of the Penrose triangle
(with the perfectly anticorrelated variables), and the even ones represent
the corresponding realizable bars (with perfectly correlated variables).
This ordering highlights the fact that the system is composed of three
disjoint $2\times2$ subsystems (formally, cyclic systems of rank
2). We call this way of representing the impossible figure \emph{Option
1}.

Another way (\emph{Option 2}) to approach the construction of a system
of epistemic random variables describing the Penrose triangle is to
look at the sequence of the labels for the cuts in the entire triangle.
We arbitrarily choose one of the six cuts in the figure as a starting
point, and then proceed to the second cut in the same bar, then to
the nearest cut in the adjacent bar, etc. This produces one of the
two patterns shown below, representing three starting points each: 
\begin{center}
\begin{equation}
\begin{array}{|c|c|c|c|c|c||c|}
\hline 1 & 0 & 1 & 0 & 1 & 0 & c_{1}\\
\hline\hline q_{1} & q_{2} & q_{3} & q_{4} & q_{5} & q_{6} & 
\\\hline \end{array}\textnormal{ or }\begin{array}{|c|c|c|c|c|c||c|}
\hline 0 & 1 & 0 & 1 & 0 & 1 & c_{1}\\
\hline\hline q_{1} & q_{2} & q_{3} & q_{4} & q_{5} & q_{6} & 
\\\hline \end{array}\:.
\end{equation}
\par\end{center}

By uniformly mixing these patterns, we obtain a vector of six jointly
distributed epistemic random variables that represent the Penrose
triangle in a single context of random variables:

\begin{equation}
\left(R_{1}^{1},R_{2}^{1},R_{3}^{1},R_{4}^{1},R_{5}^{1},R_{6}^{1}\right)\overset{dist}{=}\begin{array}{|c|c|c|}
\hline \left(1,0,1,0,1,0\right) & \left(0,1,0,1,0,1\right) & \textnormal{any other pattern}\\
\hline\hline \nicefrac{1}{2} & \nicefrac{1}{2} & 0
\\\hline \end{array}\:.
\end{equation}
The numbers in the second row are probabilities of the values in the
first row, $\overset{dist}{=}$ stands for ``distributed as.''

To complete Option 2, we have to add for comparison the epistemic
random variables describing a realizable triangle in the same fashion.
We select the triangle depictions that may be obtained by {oblique}
projection at arbitrary angles excluding multiples of 60 deg.\footnote{We exclude the multiples of 60 deg because at these angles one of
the bars is drawn with a single visible side instead of two.} Except for rotations, this process produces two distinct figures
with respect to the patterns of $0$ and $1$ that describe them in
a similar way to the Penrose triangle above. The three left patterns
in (\ref{eq:opt2RealizableTriangle}) below are the possible patterns
that describe the triangle in Figure \ref{fig:Tribar-at-30}, and
the three patterns on the right describe the one in Figure \ref{fig:Tribar-at-90}.

\begin{equation}
\begin{array}{rc}
\begin{array}{|c|c|c|c|c|c||c|}
\hline 0 & 0 & 0 & 0 & 1 & 1 & c_{2}\\
\hline\hline q_{1} & q_{2} & q_{3} & q_{4} & q_{5} & q_{6} & 
\\\hline \end{array}\textnormal{} & \begin{array}{|c|c|c|c|c|c||c|}
\hline 1 & 1 & 1 & 1 & 0 & 0 & c_{2}\\
\hline\hline q_{1} & q_{2} & q_{3} & q_{4} & q_{5} & q_{6} & 
\\\hline \end{array}\\
\\
\begin{array}{|c|c|c|c|c|c||c|}
\hline 0 & 0 & 1 & 1 & 0 & 0 & c_{2}\\
\hline\hline q_{1} & q_{2} & q_{3} & q_{4} & q_{5} & q_{6} & 
\\\hline \end{array}\textnormal{} & \begin{array}{|c|c|c|c|c|c||c|}
\hline 1 & 1 & 0 & 0 & 1 & 1 & c_{2}\\
\hline\hline q_{1} & q_{2} & q_{3} & q_{4} & q_{5} & q_{6} & 
\\\hline \end{array}\\
\\
\begin{array}{|c|c|c|c|c|c||c|}
\hline 1 & 1 & 0 & 0 & 0 & 0 & c_{2}\\
\hline\hline q_{1} & q_{2} & q_{3} & q_{4} & q_{5} & q_{6} & 
\\\hline \end{array}\textnormal{} & \begin{array}{|c|c|c|c|c|c||c|}
\hline 0 & 0 & 1 & 1 & 1 & 1 & c_{2}\\
\hline\hline q_{1} & q_{2} & q_{3} & q_{4} & q_{5} & q_{6} & 
\\\hline \end{array}
\end{array}\:.\label{eq:opt2RealizableTriangle}
\end{equation}
By taking the uniform mixture of these deterministic patterns, we
produce a joint distribution that represents a realizable triangle
as the second context in the context-content matrix $\mathcal{\mathcal{P}}_{3}^{2}$
below 
\begin{center}
\text{\emph{Option 2}} 
\begin{align}
\begin{array}{|c|c|c|c|c|c||c|}
\hline R_{1}^{1} & R_{2}^{1} & R_{3}^{1} & R_{4}^{1} & R_{5}^{1} & R_{6}^{1} & c_{1}=\text{Penrose triangle}\\
\hline R_{1}^{2} & R_{2}^{2} & R_{3}^{2} & R_{4}^{2} & R_{5}^{2} & R_{6}^{2} & c_{2}=\text{Regular triangle}\\
\hline\hline q_{1} & q_{2} & q_{3} & q_{4} & q_{5} & q_{6} & \mathcal{P}_{3}^{2}
\\\hline \end{array}.\label{eq:triangleOption2}
\end{align}
\par\end{center}

Whichever of the two options we choose, the pairs of random variables
that represent two cuts on the same bar are perfectly negatively correlated
in the case of the Penrose triangle, and are perfectly positively
correlated in a realizable triangle. Because of this the systems in
both Options 1 and 2 are contextual. In both cases the contextuality
is achieved at level 2. Therefore their hierarchical representation
is a one-component vector. The values are $\text{CNT}_{2}^{2}=1.5$
in Option 1 and $\text{CNT}_{2}^{2}=4.5$ in Option 2.\footnote{At this introductory stage we have not discussed the issue of normalization
of the (non)contextuality values, because of which we should avoid
comparing the values computed for systems of different format (such
as our Option 1 and Option 2 systems).}

The value of $\text{CNT}_{2}^{2}$ for Option 1 is clearly the sum
of the contextuality values of the three disjoint subsystems. Generally,
for a system $\mathcal{R}$ which is composed of $N$ disjoint systems
$\mathcal{R}_{i}$, the noncontextuality polytope

\begin{equation}
\mathbb{K=}\prod_{i=1}^{N}\mathbb{K}_{i},
\end{equation}
the Cartesian product of the corresponding noncontextuality polytopes
$\mathbb{K}_{i}$ of each system. Consequently,

\begin{equation}
\text{CNT}_{2}^{s^{*}}(\mathcal{R})=\sum_{i=1}^{N}\text{CNT}_{2}^{s^{*}}(\mathcal{R}_{i}),
\end{equation}
and

\begin{equation}
\text{NCNT}_{2}^{s}(\mathcal{R})=\min_{i=1}^{N}\{\text{NCNT}_{2}^{s}(\mathcal{R}_{i})\},
\end{equation}
for $2\leq s\leq s^{*}$.

\begin{figure}[tb]
\centering{}\begin{tikzpicture}[scale = 0.55, line cap=round,line join=round,>=triangle 45,x=1cm,y=1cm]


\fill[line width=2pt,fill=black,fill opacity=0.2] (-1, 1.73) -- (-2, 1.73) -- (0.5, 6.06) -- (3, 1.73) -- (2, 1.73) -- (0.5, 4.33) -- cycle;
\fill[line width=2pt,fill=black,fill opacity=0.2] (3, 1.73) -- (3.5, 0.87) -- (-3.5, 0.87) -- (-0.5, 6.06) -- (0.5, 6.06) -- (-2, 1.73) -- cycle;

\fill[line width=2pt,fill=black,fill opacity=0.3] (2, 1.73) -- (1, 1.73) -- (0, 3.44) -- (0.5, 4.33) -- cycle;

\fill[line width=2pt,fill=black,fill opacity=0.05] (-3.5, 0.87) -- (-3, 0) -- (3, 0) -- (3.5, 0.87) -- cycle;


\draw [line width=0.6pt] (-0.5, 6.06) -- (0.5,  6.06);
\draw [line width=0.6pt] (-2,   1.73) -- (0.5,  6.06);
\draw [line width=0.6pt] (-3.5, 0.87) -- (-0.5, 6.06);
\draw [line width=0.6pt] (0.5,  4.33) -- (-1,   1.73);

\draw [line width=0.6pt] (0.5, 4.33) -- (2,   1.73);
\draw [line width=0.6pt] (0.5, 6.06) -- (3.5, 0.87);
\draw [line width=0.6pt] (1,   1.73) -- (0,   3.44);
\draw [line width=0.6pt] (3,   0)    -- (3.5, 0.87);

\draw [line width=0.6pt] (-2,   1.73) -- (2,    1.73);
\draw [line width=0.6pt] (-3,   0)    -- (3,    0);
\draw [line width=0.6pt] (-3.5, 0.87) -- (-3,   0);
\draw [line width=0.6pt] (3.5,  0.87) -- (-3.5, 0.87);


\draw [line width=0.6pt, color=blue]         (-1, 5.20) -- (0,     5.20);
\draw (-1.3, 5.20) node {\small{0}};
\draw [line width=0.6pt, color=blue]         (-2, 1.73) -- (-3,    1.73);
\draw (-3.3, 1.73) node {\small{1}};
\draw [line width=0.6pt, color=blue]         (0,  5.20) -- (0.5,   4.33);
\draw [line width=0.6pt, dotted, color=blue] (-2, 1.73) -- (-1.45, 0.87);

\draw [line width=0.6pt, color=yellow]         (3,   1.73) -- (2,    1.73);
\draw (3.3, 1.73) node {\small{0}};
\draw [line width=0.6pt, dotted, color=yellow] (2,   1.73) -- (1.5,  0.87);
\draw [line width=0.6pt, color=yellow]         (0.5, 4.33) -- (1,    5.20);
\draw (1.3, 5.20) node {\small{0}};
\draw [line width=0.6pt, dotted, color=yellow] (0.5, 4.33) -- (-0.5, 4.33);

\draw [line width=0.6pt, color=red] (-2,   1.73) -- (-2.5, 0.87);
\draw [line width=0.6pt, color=red] (-2.5, 0.87) -- (-2,   0);
\draw (-2, -0.3) node {\small{1}};
\draw [line width=0.6pt, color=red] (2,    1.73) -- (2.5,  0.87);
\draw [line width=0.6pt, color=red] (2,    0)    -- (2.5,  0.87);
\draw (2, -0.3) node {\small{1}};

\end{tikzpicture} \caption{\label{fig:Alternative-impossible-tribar}Alternative impossible triangle
with cross-sections {and corresponding labels (0 or 1).} }
\end{figure}
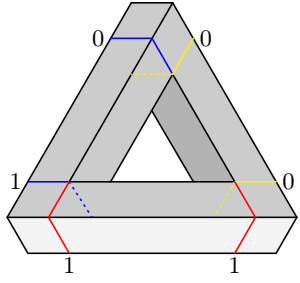

\section{Other impossible figures}

To further explore how our Option 1 and Option 2 representations capture
the impossibility of a figure by the degree of contextuality of the
resulting systems, let us consider an alternative impossible triangle,
depicted in Figure \ref{fig:Alternative-impossible-tribar}. Unlike
the Penrose triangle, this one has only one bar (the left one) with
different labels at its ends. For the bottom bar, both ends are coded
as $1$, and for the right bar the two ends are coded as $0$. Intuitively,
this triangle seems ``less impossible'' than the Penrose one. Following
the same Option 1 procedure as before, with the same variables describing
a realizable triangle, we obtain the system of the same format as
system $\mathcal{P}_{3}^{1}$; however, in two of the disjoint $2\times2$
subsystems, in the odd-numbered contexts representing the bottom and
right bars, the two random variables become deterministic, making
thereby these subsystems noncontextual.\footnote{In CbD, any deterministic variable in a system can be deleted from
the system without affecting its (non)contextuality, and the same
is true for any variables that is alone in its column (connection).
Therefore, two of the subsystems, corresponding to the lower and the
right bars, can be removed without affecting the analysis.} For Option 2, we obtain a system of the same format as $\mathcal{P}_{3}^{2}$,
but the distribution of the bunch corresponding to $c_{1}$ (impossible
triangle) is given by the uniform mixture of the patterns below:

\begin{equation}
\begin{array}{rc}
\begin{array}{|c|c|c|c|c|c||c|}
\hline 0 & 0 & 1 & 1 & 0 & 1 & c_{1}\\
\hline\hline q_{1} & q_{2} & q_{3} & q_{4} & q_{5} & q_{6} & 
\\\hline \end{array}\textnormal{} & \begin{array}{|c|c|c|c|c|c||c|}
\hline 1 & 1 & 0 & 0 & 1 & 0 & c_{1}\\
\hline\hline q_{1} & q_{2} & q_{3} & q_{4} & q_{5} & q_{6} & 
\\\hline \end{array}\\
\\
\begin{array}{|c|c|c|c|c|c||c|}
\hline 0 & 0 & 1 & 0 & 1 & 1 & c_{1}\\
\hline\hline q_{1} & q_{2} & q_{3} & q_{4} & q_{5} & q_{6} & 
\\\hline \end{array}\textnormal{} & \begin{array}{|c|c|c|c|c|c||c|}
\hline 1 & 1 & 0 & 1 & 0 & 0 & c_{1}\\
\hline\hline q_{1} & q_{2} & q_{3} & q_{4} & q_{5} & q_{6} & 
\\\hline \end{array}\\
\\
\begin{array}{|c|c|c|c|c|c||c|}
\hline 0 & 1 & 0 & 0 & 1 & 1 & c_{1}\\
\hline\hline q_{1} & q_{2} & q_{3} & q_{4} & q_{5} & q_{6} & 
\\\hline \end{array}\textnormal{} & \begin{array}{|c|c|c|c|c|c||c|}
\hline 1 & 0 & 1 & 1 & 0 & 0 & c_{1}\\
\hline\hline q_{1} & q_{2} & q_{3} & q_{4} & q_{5} & q_{6} & 
\\\hline \end{array}
\end{array}\label{eq:opt2AlternativeTriangle}
\end{equation}
{The deterministic patterns whose mixture produces random variables
variables in context $c_{2}$ are the same as in (\ref{eq:opt2RealizableTriangle}).}
The alternative impossible triangle is contextual under both representations,
with $\text{CNT}_{2}^{2}=0.5$ under Option 1, and $\text{CNT}_{2}^{2}=1.5$
under Option 2, both values being lower than the corresponding ones
for the Penrose triangle.

Finally, we need to check that the procedures above yield no contextuality
if an impossible figure is replaced with a realizable one. We use
the realizable triangle in Figure \ref{fig:Tribar-at-30}. Option
1 produces three deterministic pairs of random variables for all three
odd-numbered contexts corresponding to the pictured triangle. For
Option 2, the realizable triangle is represented by a context with
the joint distribution given by uniform mixture of the following patterns:

\begin{equation}
\begin{array}{r}
\begin{array}{|c|c|c|c|c|c||c|}
\hline 1 & 1 & 0 & 0 & 0 & 0 & c_{1}\\
\hline\hline q_{1} & q_{2} & q_{3} & q_{4} & q_{5} & q_{6} & 
\\\hline \end{array}\\
\\
\begin{array}{|c|c|c|c|c|c||c|}
\hline 0 & 0 & 1 & 1 & 0 & 0 & c_{1}\\
\hline\hline q_{1} & q_{2} & q_{3} & q_{4} & q_{5} & q_{6} & 
\\\hline \end{array}\\
\\
\begin{array}{|c|c|c|c|c|c||c|}
\hline 0 & 0 & 0 & 0 & 1 & 1 & c_{1}\\
\hline\hline q_{1} & q_{2} & q_{3} & q_{4} & q_{5} & q_{6} & 
\\\hline \end{array}
\end{array}\label{eq:opt2RealizableTriangle_c1}
\end{equation}
The realizable triangle is noncontextual under both options, and all
its hierarchical $\text{NCNT}_{2}^{s}$ measures are zero (indicating
that the systems for the realizable triangle lie on the surface of
the corresponding noncontextuality polytopes).

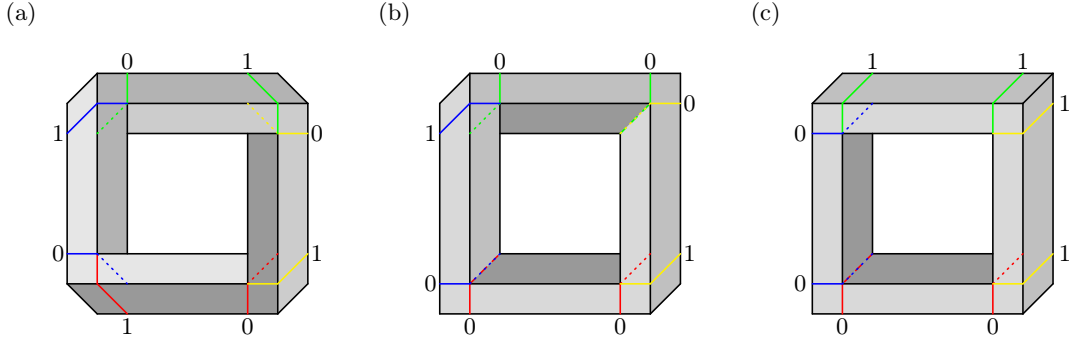
\begin{figure}[tb]
\begin{subfigure}[t]{0.32\textwidth} \caption{\raggedright\label{fig:Impossible-square.}}

\begin{centering}
\begin{tikzpicture}[scale = 0.4, line cap=round,line join=round,>=triangle 45,x=1cm,y=1cm]
\fill[line width=0.6pt,fill=black,fill opacity=0.4] (0,1) -- (1,0) -- (7,0) -- (7,6) -- (6,6) -- (6,1) -- cycle;
\fill[line width=0.6pt,fill=black,fill opacity=0.2] (7,0) -- (8,1) -- (8,7) -- (2,7) -- (2,6) -- (7,6) -- cycle;
\fill[line width=0.6pt,fill=black,fill opacity=0.3] (8,7) -- (7,8) -- (1,8) -- (1,2) -- (2,2) -- (2,7) -- cycle;
\fill[line width=0.6pt,fill=black,fill opacity=0.1] (1,8) -- (0,7) -- (0,1) -- (6,1) -- (6,2) -- (1,2) -- cycle;

\draw [line width=0.6pt] (0,1)-- (1,0);
\draw [line width=0.6pt] (1,0)-- (7,0);
\draw [line width=0.6pt] (7,0)-- (7,6);
\draw [line width=0.6pt] (6,6)-- (6,1);
\draw [line width=0.6pt] (6,1)-- (0,1);
\draw [line width=0.6pt] (7,0)-- (8,1);
\draw [line width=0.6pt] (8,1)-- (8,7);
\draw [line width=0.6pt] (8,7)-- (2,7);
\draw [line width=0.6pt] (2,6)-- (7,6);
\draw [line width=0.6pt] (8,7)-- (7,8);
\draw [line width=0.6pt] (7,8)-- (1,8);
\draw [line width=0.6pt] (1,8)-- (1,2);
\draw [line width=0.6pt] (2,2)-- (2,7);
\draw [line width=0.6pt] (1,8)-- (0,7);
\draw [line width=0.6pt] (0,7)-- (0,1);
\draw [line width=0.6pt] (6,2)-- (1,2);

\draw [line width=0.6pt,color=red] (2,0)-- (1,1);
\draw (2, -0.4) node {\small{1}};
\draw [line width=0.6pt,color=red] (1,1)-- (1,2);
\draw [line width=0.6pt,color=red] (6,0)-- (6,1);
\draw (6, -0.4) node {\small{0}};
\draw [line width=0.6pt,dotted,color=red] (6,1)-- (7,2);

\draw [line width=0.6pt,color=yellow] (6,1)-- (7,1);
\draw [line width=0.6pt,color=yellow] (7,1)-- (8,2);
\draw (8.3, 2) node {\small{1}};
\draw [line width=0.6pt,color=yellow] (7,6)-- (8,6);
\draw (8.3, 6) node {\small{0}};
\draw [line width=0.6pt,dotted,color=yellow] (7,6)-- (6,7);

\draw [line width=0.6pt,color=green] (7,6)-- (7,7);
\draw [line width=0.6pt,color=green] (7,7)-- (6,8);
\draw (6, 8.4) node {\small{1}};
\draw [line width=0.6pt,color=green] (2,8)-- (2,7);
\draw (2, 8.4) node {\small{0}};
\draw [line width=0.6pt,dotted,color=green] (2,7)-- (1,6);

\draw [line width=0.6pt,color=blue] (1,7)-- (2,7);
\draw [line width=0.6pt,color=blue] (0,6)-- (1,7);
\draw (-0.3, 6) node {\small{1}};
\draw [line width=0.6pt,color=blue] (0,2)-- (1,2);
\draw (-0.3, 2) node {\small{0}};
\draw [line width=0.6pt,dotted,color=blue] (1,2)-- (2,1);
\end{tikzpicture} 
\par\end{centering}
\end{subfigure} \begin{subfigure}[t]{0.32\textwidth} \caption{\raggedright\label{fig:Alternative-impossible-square.}}

\begin{centering}
\begin{tikzpicture}[scale = 0.4, line cap=round,line join=round,>=triangle 45,x=1cm,y=1cm]

\fill[line width=0.6pt, fill=black, fill opacity=0.15] (1, 8) -- (0, 7) -- (0, 0) -- (7, 0) -- (7, 7) -- (6, 6) -- (6, 1) -- (1, 1) -- cycle;
\fill[line width=0.6pt, fill=black, fill opacity=0.25] (7, 0) -- (8, 1) -- (8, 8) -- (1, 8) -- (1, 1) -- (2, 2) -- (2, 7) -- (7, 7) -- cycle;
\fill[line width=0.6pt, fill=black, fill opacity=0.4] (1, 1) -- (2, 2) -- (6, 2) -- (6, 1) -- cycle;
\fill[line width=0.6pt, fill=black, fill opacity=0.4] (6, 6) -- (2, 6) -- (2, 7) -- (7, 7) -- cycle;


\draw [line width=0.6pt] (1, 8) -- (0, 7);
\draw [line width=0.6pt] (0, 7) -- (0, 0);
\draw [line width=0.6pt] (0, 0) -- (7, 0);
\draw [line width=0.6pt] (7, 0) -- (7, 7);
\draw [line width=0.6pt] (6, 6) -- (6, 1);
\draw [line width=0.6pt] (6, 1) -- (1, 1);
\draw [line width=0.6pt] (1, 1) -- (1, 8);
\draw [line width=0.6pt] (7, 0) -- (8, 1);
\draw [line width=0.6pt] (8, 1) -- (8, 8);
\draw [line width=0.6pt] (8, 8) -- (1, 8);
\draw [line width=0.6pt] (2, 2) -- (2, 7);
\draw [line width=0.6pt] (2, 7) -- (7, 7);
\draw [line width=0.6pt] (2, 2) -- (6, 2);
\draw [line width=0.6pt] (6, 6) -- (2, 6);


\draw [line width=0.6pt, color=red]         (1, 0) -- (1, 1);
\draw (1, -0.4) node {\small{0}};
\draw [line width=0.6pt, dashed, color=red] (1, 1) -- (2, 2);
\draw [line width=0.6pt, color=red]         (6, 0) -- (6, 1);
\draw (6, -0.4) node {\small{0}};
\draw [line width=0.6pt, dotted, color=red] (6, 1) -- (7, 2);

\draw [line width=0.6pt, dashed, color=green] (6, 6) -- (7, 7);
\draw [line width=0.6pt, color=green]         (7, 7) -- (7, 8);
\draw (7, 8.4) node {\small{0}};
\draw [line width=0.6pt, color=green]         (2, 8) -- (2, 7);
\draw (2, 8.4) node {\small{0}};
\draw [line width=0.6pt, dotted, color=green] (2, 7) -- (1, 6);

\draw [line width=0.6pt, color=yellow]         (7, 1) -- (8, 2);
\draw (8.3, 2) node {\small{1}};
\draw [line width=0.6pt, color=yellow]         (7, 7) -- (8, 7);
\draw (8.3, 7) node {\small{0}};
\draw [line width=0.6pt, color=yellow]         (6, 1) -- (7, 1);
\draw [line width=0.6pt, dotted, color=yellow] (6, 6) -- (7, 7);

\draw [line width=0.6pt, color=blue]         (1, 7) -- (2, 7);
\draw [line width=0.6pt, color=blue]         (0, 6) -- (1, 7);
\draw (-0.3, 6) node {\small{1}};
\draw [line width=0.6pt, color=blue]         (1, 1) -- (0, 1);
\draw (-0.3, 1) node {\small{0}};
\draw [line width=0.6pt, dotted, color=blue] (1, 1) -- (2, 2);
\end{tikzpicture} 
\par\end{centering}
\end{subfigure} \begin{subfigure}[t]{0.32\textwidth} \caption{\raggedright\label{fig:Realizable-square.}}

\begin{centering}
\begin{tikzpicture}[scale = 0.4, rotate = 180, line cap=round,line join=round,>=triangle 45,x=1cm,y=1cm]

\fill[line width=0.6pt, fill=black, fill opacity=0.25] (0, 0) -- (7, 0) -- (8, 1) -- (1, 1) -- cycle;
\fill[line width=0.6pt, fill=black, fill opacity=0.25] (0, 0) -- (1, 1) -- (1, 8) -- (0, 7) -- cycle;
\fill[line width=0.6pt, fill=black, fill opacity=0.15] (1, 1) -- (8, 1) -- (8, 8) -- (1, 8) -- cycle;
\fill[line width=0.6pt, fill=white, fill opacity=1.0] (2, 2) -- (7, 2) -- (7, 7) -- (2, 7) -- cycle;
\fill[line width=0.6pt, fill=black, fill opacity=0.4] (2, 7) -- (7, 7) -- (6, 6) -- (2, 6) -- cycle;
\fill[line width=0.6pt, fill=black, fill opacity=0.4] (7, 2) -- (6, 2) -- (6, 6) -- (7, 7) -- cycle;


\draw [line width=0.6pt] (0, 0) -- (1, 1);
\draw [line width=0.6pt] (0, 0) -- (7, 0);
\draw [line width=0.6pt] (0, 7) -- (0, 0);
\draw [line width=0.6pt] (1, 1) -- (1, 8);
\draw [line width=0.6pt] (1, 1) -- (8, 1);
\draw [line width=0.6pt] (1, 8) -- (0, 7);
\draw [line width=0.6pt] (2, 2) -- (7, 2);
\draw [line width=0.6pt] (2, 7) -- (2, 2);
\draw [line width=0.6pt] (2, 7) -- (7, 7);
\draw [line width=0.6pt] (6, 2) -- (6, 6);
\draw [line width=0.6pt] (6, 6) -- (2, 6);
\draw [line width=0.6pt] (7, 0) -- (8, 1);
\draw [line width=0.6pt] (7, 2) -- (7, 7);
\draw [line width=0.6pt] (8, 1) -- (8, 8);
\draw [line width=0.6pt] (8, 8) -- (1, 8);


\draw [line width=0.6pt, color=green] (1, 0) -- (2, 1);
\draw (1, -0.4) node {\small{1}};
\draw [line width=0.6pt, color=green] (2, 1) -- (2, 2);
\draw [line width=0.6pt, color=green] (6, 0) -- (7, 1);
\draw (6, -0.4) node {\small{1}};
\draw [line width=0.6pt, color=green] (7, 1) -- (7, 2);

\draw [line width=0.6pt, dashed, color=red] (6, 6) -- (7, 7);
\draw [line width=0.6pt, color=red]         (7, 7) -- (7, 8);
\draw (7, 8.4) node {\small{0}};
\draw [line width=0.6pt, color=red]         (2, 8) -- (2, 7);
\draw (2, 8.4) node {\small{0}};
\draw [line width=0.6pt, dotted, color=red] (2, 7) -- (1, 6);

\draw [line width=0.6pt, color=blue]         (7, 2) -- (8, 2);
\draw (8.4, 2) node {\small{0}};
\draw [line width=0.6pt, color=blue]         (7, 7) -- (8, 7);
\draw (8.4, 7) node {\small{0}};
\draw [line width=0.6pt, dotted, color=blue] (6, 1) -- (7, 2);
\draw [line width=0.6pt, dotted, color=blue] (6, 6) -- (7, 7);

\draw [line width=0.6pt, color=yellow] (1, 7) -- (2, 7);
\draw [line width=0.6pt, color=yellow] (0, 6) -- (1, 7);
\draw (-0.4, 6) node {\small{1}};
\draw [line width=0.6pt, color=yellow] (1, 2) -- (2, 2);
\draw [line width=0.6pt, color=yellow] (1, 2) -- (0, 1);
\draw (-0.4, 1) node {\small{1}};
\end{tikzpicture} 
\par\end{centering}
\end{subfigure} \caption{\label{fig:Square-figures.}Square figures{, shown with cuts and
corresponding labels (0 or 1).} (\textbf{a}) Impossible square. (\textbf{b})
Alternative impossible square. (\textbf{c}) Realizable ``normal''
square. }
\end{figure}

Contextuality analysis similar to the reported above for the impossible
triangles can be extended to other impossible figures. We have conducted
this analysis for the impossible square (or rectangle) and the impossible
circle (also known as impossible loop). For the impossible square,
in addition to the figure constructed with the same type of corners
as in the Penrose triangle (Figure \ref{fig:Impossible-square.}),
we have considered an alternative impossible square (Figure \ref{fig:Alternative-impossible-square.}).
Figure \ref{fig:Realizable-square.} shows a realizable (``normal'')
square. Since the procedures and reasoning here are in all essential
details the same as for the impossible triangles, we have relegated
the details to Appendix A.

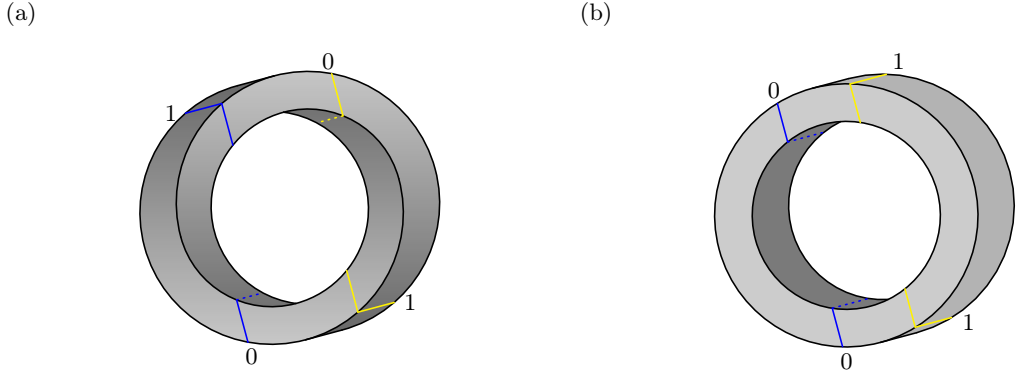
\begin{figure}
\begin{subfigure}[t]{0.5\textwidth} \caption{\raggedright\label{fig:Impossible-circle.}}
\centering \begin{tikzpicture}[rotate=15, scale = 0.5, line cap=round,line join=round,>=triangle 45,x=1cm,y=1cm]


\shade[bottom color=black!20, top color=black!60]
(-0.5, 3.5)  arc (90:270:3.5) -- 
(-0.5, -3.5) arc (270:360:3.5) --
(3,    0)    arc (0:90:2.5) --
(0.5,  2.5)  arc (67.3801:-90:2.6) --
(-0.5, -2.5) arc (270:180:2.5) --
(-3,   0)    arc (180:90:3.5)
-- cycle;

\shade[bottom color=black!60, top color=black!20]
(0.5, -3.5)  arc (-90:90:3.5) -- 
(0.5, 3.5)   arc (90:180:3.5) --
(-3,  0)     arc (180:270:2.5) --
(-0.5, -2.5) arc (180+67.3801:90:2.6) --
(0.5, 2.5)   arc (90:0:2.5) --
(3,   0)     arc (0:-90:3.5)
-- cycle;


\draw [line width=0.6pt] (0.5, 3.5)  -- (-0.5, 3.5) ;
\draw [line width=0.6pt]
(-0.5, 3.5)  arc (90:270:3.5) ;
\draw [line width=0.6pt]
(-0.5, -3.5) arc (270:360:3.5) ;
\draw [line width=0.6pt]
(3,    0)    arc (0:90:2.5) ;
\draw [line width=0.6pt]
(0.5,  2.5)  arc (67.3801:-90:2.6) ;
\draw [line width=0.6pt]
(-0.5, -2.5) arc (270:180:2.5) ;
\draw [line width=0.6pt]
(-3,   0)    arc (180:90:3.5) ;

\draw [line width=0.6pt] (0.5, -3.5) -- (-0.5, -3.5) ;
\draw [line width=0.6pt]
(0.5, -3.5)  arc (-90:90:3.5) ; 
\draw [line width=0.6pt]
(-0.5, -2.5) arc (180+67.3801:90:2.6) ;


\draw [line width=0.6pt, color=blue]        (-1.03, 3.15) -- (-2.03, 3.15);
\draw (-2.4, 3.25) node {\small{1}};
\draw [line width=0.6pt, color=blue]        (-1.03, 2) -- (-1.03, 3.15);
\draw [line width=0.6pt, color=blue]        (-2,-2) -- (-2, -3.17);
\draw (-2, -3.55) node {\small{0}};
\draw [line width=0.6pt,dotted, color=blue] (-2,-2) -- (-1.27, -2);

\draw [line width=0.6pt, color=yellow]        (1.03, -3.15) -- (2.03, -3.15);
\draw (2.45, -3.25) node {\small{1}};
\draw [line width=0.6pt, color=yellow]        (2, 2) -- (2, 3.16);
\draw (2, 3.5) node {\small{0}};
\draw [line width=0.6pt, color=yellow]        (1.03, -2) -- (1.03, -3.15);
\draw [line width=0.6pt,dotted, color=yellow] (2, 2) -- (1.27, 2);

\end{tikzpicture}

\end{subfigure} \begin{subfigure}[t]{0.5\textwidth} \caption{\raggedright\label{fig:Realizable-circle.}}

\begin{centering}
\begin{tikzpicture}[rotate = 15, scale = 0.5, line cap=round,line join=round,>=triangle 45,x=1cm,y=1cm]


\fill [fill=black, fill opacity = 0.2] (0,0) circle (3.5cm);
\fill [fill=black, fill opacity = 0.4] (0,0) circle (2.5cm);

\fill [fill=white] 
(0.5,  2.45) arc(78.463:-78.463:2.5) -- 
(0.5, -2.45) arc(258.463:180-78.463:2.5) ; 


\fill[fill=black, opacity=0.3]
(0, 3.5) arc(90:-90:3.5) -- 
(0, -3.5) -- (1, -3.5) --
(1, -3.5) arc(-90:90:3.5) -- 
(1, 3.5) -- (0, 3.5) ;


\draw [line width=0.6pt] (0,0) circle (2.5cm);
\draw [line width=0.6pt] (0,0) circle (3.5cm);

\draw [shift={(1,0)},line width=0.6pt]  plot[domain=-1.5707963267948966:1.5707963267948966,variable=\t]({1*3.5*cos(\t r)+0*3.5*sin(\t r)},{0*3.5*cos(\t r)+1*3.5*sin(\t r)});
\draw [line width=0.6pt] (0,-3.5)-- (1,-3.5);
\draw [line width=0.6pt] (0,3.5)-- (1,3.5);
\draw [shift={(1,0)},line width=0.6pt]  plot[domain=1.7721542475852274:4.511031059594359,variable=\t]({1*2.5*cos(\t r)+0*2.5*sin(\t r)},{0*2.5*cos(\t r)+1*2.5*sin(\t r)});


\draw [line width=0.6pt, color = yellow] (1, -3.35) -- (1, -2.29);
\draw [line width=0.6pt, color = yellow] (1, -3.35) -- (2, -3.35);
\draw (2.4, -3.6) node {\small{1}};
\draw [line width=0.6pt, color = yellow] (1,  3.35) -- (1,  2.29);
\draw [line width=0.6pt, color = yellow] (1,  3.35) -- (2,  3.35);
\draw (2.4, 3.6) node {\small{1}};

\draw [line width=0.6pt, color = blue]         (-1, -3.35) -- (-1, -2.29);
\draw (-1, 3.7) node {\small{0}};
\draw [line width=0.6pt, color = blue]         (-1,  3.35) -- (-1, 2.29);
\draw (-1, -3.75) node {\small{0}};
\draw [line width=0.6pt, dotted, color = blue] (-1,  2.29) -- (0,  2.29);
\draw [line width=0.6pt, dotted, color = blue] (-1, -2.29) -- (0, -2.29);

\end{tikzpicture} 
\par\end{centering}
\end{subfigure} \caption{\label{fig:Circle-figures.}Circular figures{, shown with cuts and
corresponding labels (0 or 1).} (\textbf{a}){\,}Impossible circle.
(\textbf{b}){\,}Realizable circle. }
\end{figure}

Figure \ref{fig:Circle-figures.} depicts an impossible circle (\ref{fig:Impossible-circle.})
and a realizable one (\ref{fig:Realizable-circle.}). To characterize
the circles, we may look at them as being composed of two handles
joined by curved bars, with the cut lines drawn between the handles
and the joining bars. In this way, we obtain systems analogous to
those we found for the two ways we are using to represent the figures.
Again, we relegate the details of the analysis to Appendix A.

We {summarize} the results of our analysis in Table \ref{tab:Measures-of-contextuality-ImpFig}.
As we see, all the impossible figures we explored, under both Option
1 and Option 2, are contextual at level 2.

\begin{table}[bh]
\caption{\label{tab:Measures-of-contextuality-ImpFig}Measures of contextuality
for the impossible figures}

\centering{}%
\begin{tabular}{lcc}
\hline 
Impossible figure  & $\begin{array}{c}
\textnormal{Option 1}\\
\text{CNT}_{2}^{2}
\end{array}$  & $\begin{array}{c}
\textnormal{Option 2}\\
\text{CNT}_{2}^{2}
\end{array}$\tabularnewline
\hline 
Triangle in Figure \ref{fig:Penrose-tribar.}  & $1.5$  & $4.5$\tabularnewline
Triangle in Figure \ref{fig:Alternative-impossible-tribar}  & $0.5$  & $1.5$\tabularnewline
Square in Figure \ref{fig:Impossible-square.}  & $2.0$  & $8.0$\tabularnewline
Square in Figure \ref{fig:Alternative-impossible-square.}  & $1.0$  & $2.0$\tabularnewline
Circle in Figure \ref{fig:Impossible-circle.}  & $1.0$  & $2.0$\tabularnewline
\hline 
\end{tabular}
\end{table}

\section{Escher's ``Ascending and Descending''}

\label{eschers-ascending-and-descending-staircase}

\begin{figure}
\centering \includegraphics[width=0.55\textwidth]{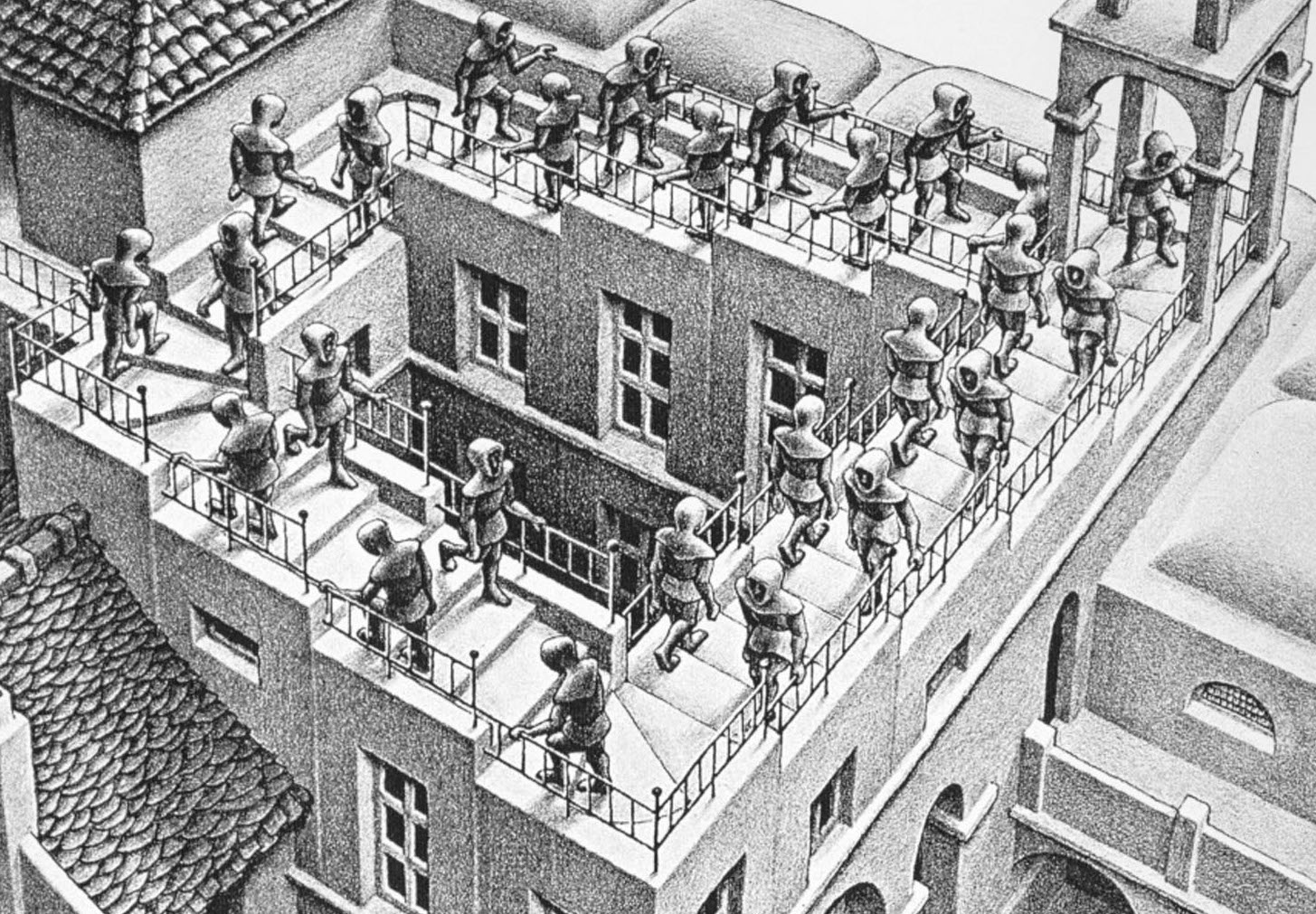}
\caption{\label{fig:Ascending-and-descending.}Fragment of M.\ C.\ Escher's
lithograph ``Ascending and Descending.''}
\end{figure}

We approach the Ascending and Descending lithograph by M. C. Escher
(Figure \ref{fig:Ascending-and-descending.}) by considering the four
stair flights as four contents, $q_{1},q_{2},q_{3},q_{4}$. The same
as for the impossible figures, we have two ways of constructing the
system of epistemic random variables. In Option 1, the system is represented
by the following context-content matrix: 
\begin{center}
\text{\emph{Option 1}} 
\begin{align}
\begin{array}{|c|c|c|c||c|}
\hline R_{1}^{1} & R_{2}^{1} &  &  & c_{1}\\
\hline  & R_{2}^{2} & R_{3}^{2} &  & c_{2}\\
\hline  &  & R_{3}^{3} & R_{4}^{3} & c_{3}\\
\hline R_{1}^{4} &  &  & R_{4}^{4} & c_{4}\\
\hline\hline R_{1}^{5} & R_{2}^{5} & R_{3}^{5} & R_{4}^{5} & c_{5}\\
\hline\hline q_{1} & q_{2} & q_{3} & q_{4} & \mathcal{E}^{1}
\\\hline \end{array}.\label{eq:EscherOption1}
\end{align}
\par\end{center}

If we view a stair flight as ascending, then $R_{q}^{c}=1$, otherwise
$R_{q}^{c}=0$. For every pair of consecutive stair flights in the
picture, they are seen as both ascending, or both descending. By uniformly
mixing

\begin{equation}
\begin{array}{|c|c||c|}
\hline 0 & 0 & c\\
\hline\hline q & q' & 
\\\hline \end{array}\textnormal{ and }\begin{array}{|c|c||c|}
\hline 1 & 1 & c\\
\hline\hline q & q' & 
\\\hline \end{array},
\end{equation}
we form the first four contexts in the system. Thus, in contexts $c_{1}$
to $c_{4}$, the distributions are described by

\begin{equation}
\langle R_{k}^{k}\rangle=\langle R_{k\oplus1}^{k}\rangle=0.5,\langle R_{k}^{k}R_{k\oplus1}^{k}\rangle=1,
\end{equation}
for $k=1,\ldots,4$ ($\oplus$ is cyclic addition, with $4\oplus1=1$).
The fifth context in $\mathcal{E}^{1}$ includes the possible patterns
of ascent and descent for the entire staircase. The strangeness (or
impossibility) of the situation in the lithograph is that four stair
flights forming a closed loop cannot ascend or descend indefinitely:
the number of ascending stair flights should be precisely two to counterbalance
the descending ones. In other words, the physically possible values
of the bunch $\left(R_{1}^{5},R_{2}^{5},R_{3}^{5},R_{4}^{5}\right)$
are

\begin{equation}
\begin{array}{rcc}
\begin{array}{|c|c|c|c||c|}
\hline 0 & 0 & 1 & 1 & c_{5}\\
\hline\hline q_{1} & q_{2} & q_{3} & q_{4} & 
\\\hline \end{array} &  & \begin{array}{|c|c|c|c||c|}
\hline 1 & 1 & 0 & 0 & c_{5}\\
\hline\hline q_{1} & q_{2} & q_{3} & q_{4} & 
\\\hline \end{array}\\
\\
\begin{array}{|c|c|c|c||c|}
\hline 0 & 1 & 0 & 1 & c_{2}\\
\hline\hline q_{1} & q_{2} & q_{3} & q_{4} & 
\\\hline \end{array} &  & \begin{array}{|c|c|c|c||c|}
\hline 1 & 0 & 1 & 0 & c_{2}\\
\hline\hline q_{1} & q_{2} & q_{3} & q_{4} & 
\\\hline \end{array}\\
\\
\begin{array}{|c|c|c|c||c|}
\hline 1 & 0 & 0 & 1 & c_{2}\\
\hline\hline q_{1} & q_{2} & q_{3} & q_{4} & 
\\\hline \end{array} &  & \begin{array}{|c|c|c|c||c|}
\hline 0 & 1 & 1 & 0 & c_{2}\\
\hline\hline q_{1} & q_{2} & q_{3} & q_{4} & 
\\\hline \end{array}
\end{array}.
\end{equation}
The epistemic distribution of the fifth bunch therefore is given by
the uniform mixture of these deterministic patterns.

The second way in which we represent the Ascending-Descending staircase
is by looking at the four stair flights together. This forms one (``impossible'')
context where either all staircases are described as ascending ($R_{q}^{1}=1,\text{ for }q=1,\ldots,4$)
or all of them are descending ($R_{q}^{1}=0,\text{ for }i=1,\ldots,4$).
A second context, describing the physically realizable patterns is
formed in the same way as the fifth context above. The system we obtain
in this way is 
\begin{center}
\emph{Option 2} 
\begin{align}
\begin{array}{|c|c|c|c||c|}
\hline R_{1}^{1} & R_{2}^{1} & R_{3}^{1} & R_{4}^{1} & c_{1}\\
\hline R_{1}^{2} & R_{2}^{2} & R_{3}^{2} & R_{4}^{2} & c_{2}\\
\hline\hline q_{1} & q_{2} & q_{3} & q_{4} & \mathcal{E}^{2}
\\\hline \end{array}.\label{eq:EscherOption2}
\end{align}
\par\end{center}

\noindent For both these options, the resulting systems are contextual
at the level $s^{*}=2$, and the values are $\text{CNT}_{2}^{2}=\nicefrac{4}{3}$
for Option 1 and $\text{CNT}_{2}^{2}=2$ for Option 2.

\section{Conclusion}

We have introduced a hierarchical measure of (non)contextuality of
systems of random variables. It follows the same logic and is calculated
similarly to the measures of contextuality $\text{CNT}_{2}$ and noncontextuality
$\text{NCNT}_{2}$. It is clear that in cyclic systems the hierarchical
measure equals $\text{CNT}_{2}$ when the system is contextual, and
it equals $\text{NCNT}_{2}$ when the system is noncontextual. It
still remains to investigate whether some of the properties of $\text{CNT}_{2}$-$\text{NCNT}_{2}$
described in \cite{DzhKujCer_Polytope} for cyclic systems generalize
to some classes of noncyclic systems.

The analysis of the impossible figures show that the intuitive degree
of the impossibility or strangeness of those figures can be captured
through the contextuality of the systems of epistemic random variables
chosen to describe them. When the endpoint codes of a bar in a Penrose-like
figure are anticorrelated, the adjacent bars appear to twist toward
different directions. The more such ``twisted'' situations we see
in a figure, the stranger it looks, and the greater its contextuality
(if achieved at the same level).

All contextual systems constructed in this paper happen to be contextual
at level 2. This cannot be otherwise for the Option 1 systems which
only contain two random variables in each bunch. However, it is an
empirical rather than mathematically deducible fact for other systems.
More work is needed to find out the scope of the impossible figures
whose reasonable descriptions have the same property.


\vspace{6pt}

\paragraph{Acknowledgments.}

The authors are grateful to Janne V. Kujala who critically analyzed
and corrected a mistake in the earlier version of the hierarchical
contextuality measure presented in this paper.

\appendix

\section{Appendix: Impossible squares and circles}

\global\long\def\theequation{\thesection.\arabic{equation}}%
 \setcounter{equation}{0}

To create the epistemic random variables depicting realizable (``normal'')
square, we proceed analogously to the realizable triangles: we systematically
consider all depictions that may be obtained by {oblique} projection
at arbitrary angles excluding the multiples of 90 deg. In this case,
ignoring rotations, there is only one pattern of endpoint labels produced
by all realizable squares. Varying the starting point, we get the
following sequences whose uniform mixture yields the epistemic random
variables for realizable squares:

\begin{equation}
\begin{array}{rc}
\begin{array}{|c|c|c|c|c|c|c|c||c|}
\hline 0 & 0 & 1 & 1 & 1 & 1 & 0 & 0 & c_{2}\\
\hline\hline q_{1} & q_{2} & q_{3} & q_{4} & q_{5} & q_{6} & q_{7} & q_{8} & 
\\\hline \end{array}\textnormal{} & \begin{array}{|c|c|c|c|c|c|c|c||c|}
\hline 1 & 1 & 0 & 0 & 0 & 0 & 1 & 1 & c_{2}\\
\hline\hline q_{1} & q_{2} & q_{3} & q_{4} & q_{5} & q_{6} & q_{7} & q_{8} & 
\\\hline \end{array}\\
\\
\begin{array}{|c|c|c|c|c|c|c|c||c|}
\hline 0 & 0 & 0 & 0 & 1 & 1 & 1 & 1 & c_{2}\\
\hline\hline q_{1} & q_{2} & q_{3} & q_{4} & q_{5} & q_{6} & q_{7} & q_{8} & 
\\\hline \end{array}\textnormal{} & \begin{array}{|c|c|c|c|c|c|c|c||c|}
\hline 1 & 1 & 1 & 1 & 0 & 0 & 0 & 0 & c_{2}\\
\hline\hline q_{1} & q_{2} & q_{3} & q_{4} & q_{5} & q_{6} & q_{7} & q_{8} & 
\\\hline \end{array}
\end{array}\label{eq:opt2RealizableSquare}
\end{equation}

All the results for impossible squares are obtained as a straightforward
expansion of those for impossible triangles. Following Option 1, we
obtain the systems describing the rectangles in Fig. 4 in format
\begin{center}
\text{\emph{Option 1}} 
\begin{align}
\begin{array}{|c|c|c|c|c|c|c|c||c|}
\hline R_{1}^{1} & R_{2}^{1} &  &  &  &  &  &  & c_{1}\\
\hline R_{1}^{2} & R_{2}^{2} &  &  &  &  &  &  & c_{2}\\
\hline  &  & R_{3}^{3} & R_{4}^{3} &  &  &  &  & c_{3}\\
\hline  &  & R_{3}^{4} & R_{4}^{4} &  &  &  &  & c_{4}\\
\hline  &  &  &  & R_{5}^{5} & R_{6}^{5} &  &  & c_{5}\\
\hline  &  &  &  & R_{5}^{6} & R_{6}^{6} &  &  & c_{6}\\
\hline  &  &  &  &  &  & R_{7}^{7} & R_{8}^{7} & c_{7}\\
\hline  &  &  &  &  &  & R_{7}^{8} & R_{8}^{8} & c_{8}\\
\hline\hline q_{1} & q_{2} & q_{3} & q_{4} & q_{5} & q_{6} & q_{7} & q_{8} & \mathcal{P}_{4}^{1}
\\\hline \end{array}.\label{eq:squareOption1}
\end{align}
\par\end{center}

\noindent The systems for Option 2 have the format
\begin{center}
\text{\emph{Option 2}} 
\begin{align}
\begin{array}{|c|c|c|c|c|c|c|c||c|}
\hline R_{1}^{1} & R_{2}^{1} & R_{3}^{1} & R_{4}^{1} & R_{5}^{1} & R_{6}^{1} & R_{7}^{1} & R_{8}^{1} & c_{1}\\
\hline R_{1}^{2} & R_{2}^{2} & R_{3}^{2} & R_{4}^{2} & R_{5}^{2} & R_{6}^{2} & R_{7}^{2} & R_{8}^{2} & c_{2}\\
\hline\hline q_{1} & q_{2} & q_{3} & q_{4} & q_{5} & q_{6} & q_{7} & q_{8} & \mathcal{P}_{4}^{2}
\\\hline \end{array}.\label{eq:squareOption2}
\end{align}
\par\end{center}

For the first option, just as with the triangles, all random variables
representing a realizable bar, are perfectly correlated. The random
variables representing bars of the Penrose-like square are negatively
correlated, whereas for the alternative impossible square, two of
the four odd-numbered bunches become deterministic. For the realizable
square in Figure \ref{fig:Realizable-square.}, all four odd-numbered
bunches are deterministic, making the system trivially noncontextual.

For Option 2, the patterns that generate the joint distribution for
the context describing the Penrose-like impossible square are

\begin{equation}
\begin{array}{r}
\begin{array}{|c|c|c|c|c|c|c|c||c|}
\hline 0 & 1 & 0 & 1 & 0 & 1 & 0 & 1 & c_{1}\\
\hline\hline q_{1} & q_{2} & q_{3} & q_{4} & q_{5} & q_{6} & q_{7} & q_{8} & 
\\\hline \end{array}\\
\\
\begin{array}{|c|c|c|c|c|c|c|c||c|}
\hline 1 & 0 & 1 & 0 & 1 & 0 & 1 & 0 & c_{1}\\
\hline\hline q_{1} & q_{2} & q_{3} & q_{4} & q_{5} & q_{6} & q_{7} & q_{8} & 
\\\hline \end{array}
\end{array}.\label{eq:opt2PenroseSquare}
\end{equation}
The patterns that generate the joint distribution for the context
describing the alternative impossible square are

\begin{equation}
\begin{array}{rc}
\begin{array}{|c|c|c|c|c|c|c|c||c|}
\hline 0 & 1 & 0 & 0 & 0 & 1 & 0 & 0 & c_{1}\\
\hline\hline q_{1} & q_{2} & q_{3} & q_{4} & q_{5} & q_{6} & q_{7} & q_{8} & 
\\\hline \end{array}\textnormal{} & \begin{array}{|c|c|c|c|c|c|c|c||c|}
\hline 0 & 0 & 0 & 1 & 0 & 0 & 0 & 1 & c_{1}\\
\hline\hline q_{1} & q_{2} & q_{3} & q_{4} & q_{5} & q_{6} & q_{7} & q_{8} & 
\\\hline \end{array}\\
\\
\begin{array}{|c|c|c|c|c|c|c|c||c|}
\hline 1 & 0 & 0 & 0 & 1 & 0 & 0 & 0 & c_{1}\\
\hline\hline q_{1} & q_{2} & q_{3} & q_{4} & q_{5} & q_{6} & q_{7} & q_{8} & 
\\\hline \end{array}\textnormal{} & \begin{array}{|c|c|c|c|c|c|c|c||c|}
\hline 0 & 0 & 1 & 0 & 0 & 0 & 1 & 0 & c_{1}\\
\hline\hline q_{1} & q_{2} & q_{3} & q_{4} & q_{5} & q_{6} & q_{7} & q_{8} & 
\\\hline \end{array}
\end{array}\label{eq:opt2AlternativeSquare}
\end{equation}

The measures of contextuality for the impossible square (\ref{fig:Impossible-square.})
are $\text{CNT}_{2}^{2}=2$ and $\text{CNT}_{2}^{2}=8$, under Options
1 and 2, respectively. The corresponding measures for the alternative
impossible square (\ref{fig:Alternative-impossible-square.}) are
$\text{CNT}_{2}^{2}=1$ and $\text{CNT}_{2}^{2}=2$. The realizable
square (\ref{fig:Realizable-square.}) is noncontextual and all its
noncontextuality measures are zero.

For the impossible circle, the system for Option 1 has the format 
\begin{center}
\text{\emph{Option 1}} 
\begin{align}
\begin{array}{|c|c|c|c||c|}
\hline R_{1}^{1} & R_{2}^{1} &  &  & c_{1}\\
\hline R_{1}^{2} & R_{2}^{2} &  &  & c_{2}\\
\hline  &  & R_{3}^{3} & R_{4}^{3} & c_{3}\\
\hline  &  & R_{3}^{4} & R_{4}^{4} & c_{4}\\
\hline\hline q_{1} & q_{2} & q_{3} & q_{4} & \mathcal{P}_{2}^{1}
\\\hline \end{array},\label{eq:circleOption1}
\end{align}
\par\end{center}

\noindent and the system for the Option 2 has the format 
\begin{center}
\text{\emph{Option 2}} 
\begin{align}
\begin{array}{|c|c|c|c||c|}
\hline R_{1}^{1} & R_{2}^{1} & R_{3}^{1} & R_{4}^{1} & c_{1}\\
\hline R_{1}^{2} & R_{2}^{2} & R_{3}^{2} & R_{4}^{2} & c_{2}\\
\hline\hline q_{1} & q_{2} & q_{3} & q_{4} & \mathcal{P}_{2}^{2}
\\\hline \end{array}.\label{eq:circleOption2}
\end{align}
\par\end{center}

\noindent The distributions of the random variables in each context
for Option 1 follows the same pattern as the corresponding Option
1 for the triangles and the squares. The two variables in a bunch
corresponding to the impossible loop are anticorrelated and the ones
corresponding to the realizable one are perfectly correlated. For
Option 2,

\begin{equation}
\begin{array}{r}
\begin{array}{|c|c|c|c||c|}
\hline 0 & 1 & 0 & 1 & c_{2}\\
\hline\hline q_{1} & q_{2} & q_{3} & q_{4} & 
\\\hline \end{array}\\
\\
\begin{array}{|c|c|c|c||c|}
\hline 1 & 0 & 1 & 0 & c_{2}\\
\hline\hline q_{1} & q_{2} & q_{3} & q_{4} & 
\\\hline \end{array}
\end{array}\label{eq:opt2PenroseCircle}
\end{equation}
gives the patterns for the impossible circle. The patterns that generate
the joint distribution for the bunch representing the realizable circle
are

\begin{equation}
\begin{array}{r}
\begin{array}{|c|c|c|c||c|}
\hline 0 & 0 & 1 & 1 & c_{2}\\
\hline\hline q_{1} & q_{2} & q_{3} & q_{4} & 
\\\hline \end{array}\\
\\
\begin{array}{|c|c|c|c||c|}
\hline 1 & 1 & 0 & 0 & c_{2}\\
\hline\hline q_{1} & q_{2} & q_{3} & q_{4} & 
\\\hline \end{array}
\end{array}.\label{eq:opt2RealizableCircle}
\end{equation}

The measures of contextuality for the impossible loop are $\text{CNT}_{2}^{2}=1$
and $\text{CNT}_{2}^{2}=2$, under Options 1 and 2, respectively.
The realizable figure is noncontextual and all its noncontextuality
measures are zero.

\section{Appendix: Measures of contextuality}

\global\long\def\thetable{\thesection.\arabic{table}}%
 \setcounter{table}{0}

For completeness and comparison purposes, we present the values of
the contextuality measures $\text{CNT}_{1}$, $\text{CNT}_{2}$, $\text{CNT}_{2}^{2}$,
$\text{CNT}_{3}$, and contextual fraction (CNTF), for each of the
systems used to represent the impossible figures. The measure $\text{CNT}_{2}^{2}$
is the hierarchical measure of (non)contextuality described in this
paper. The contextual fraction was introduced in \cite{Abramsky_Fraction}.
The other measures and the linear programming tasks that may be used
to compute them are presented in \cite{KujDzh_Measures}.

Table \ref{tab:Contextuality-measures-of-1} presents the contextuality
measures for the systems representing the triangles. Table \ref{tab:Contextuality-measures-of-2}
presents the contextuality measures for squares figures. Table \ref{tab:Contextuality-measures-of-3}
presents the contextuality measures for the systems representing the
two circles. Table \ref{tab:Contextuality-measures-of-4} presents
the contextuality measures of the systems representing Escher's Ascending
and Descending Staircase.

\begin{table}[h]
\centering{}\caption{\label{tab:Contextuality-measures-of-1}Contextuality measures of
systems representing the alternative impossible triangle and a realizable
triangle.}
\begin{tabular}{llrrrrr}
\hline 
Figure  & System  & $\text{CNT}_{1}$  & $\text{CNT}_{2}$  & $\text{CNT}_{2}^{2}$  & $\text{CNT}_{3}$  & $\text{CNTF}$\tabularnewline
\hline 
Penrose triangle  & Option 1  & $1.5$  & $1.5$  & $1.5$  & $1$  & $1$\tabularnewline
 & Option 2  & $1.5$  & $8.0$  & $4.5$  & $1$  & $1$\tabularnewline
\hline 
Alt. Imp.  & Option 1  & $0.5$  & $0.5$  & $0.5$  & $1$  & $1$\tabularnewline
 & Option 2  & $0.5$  & $3.0$  & $1.5$  & $\nicefrac{1}{3}$  & $1$\tabularnewline
\hline 
Realizable  & Option 1  & $0$  & $0$  & $0$  & $0$  & $0$\tabularnewline
 & Option 2  & $0$  & $0$  & $0$  & $0$  & $0$\tabularnewline
\hline 
\end{tabular}
\end{table}

\begin{table}[h]
\centering{}\caption{\label{tab:Contextuality-measures-of-2}Contextuality measures of
systems representing the square figures.}
\begin{tabular}{llrrrrr}
\hline 
Figure  & System  & $\text{CNT}_{1}$  & $\text{CNT}_{2}$  & $\text{CNT}_{2}^{2}$  & $\text{CNT}_{3}$  & $\text{CNTF}$\tabularnewline
\hline 
Penrose like  & Option 1  & $2$  & $2$  & $2$  & $1$  & $1$\tabularnewline
 & Option 2  & $2$  & $18$  & $8$  & $1$  & $1$\tabularnewline
\hline 
Alt. Imp.  & Option 1  & $1$  & $1$  & $1$  & $1$  & $1$\tabularnewline
 & Option 2  & $1$  & $2$  & $2$  & $\nicefrac{1}{2}$  & $1$\tabularnewline
\hline 
Realizable  & Option 1  & $0$  & $0$  & $0$  & $0$  & $0$\tabularnewline
 & Option 2  & $0$  & $0$  & $0$  & $0$  & $0$\tabularnewline
\hline 
\end{tabular}
\end{table}

\begin{table}[h]
\centering{}\caption{\label{tab:Contextuality-measures-of-3}Contextuality measures of
systems representing the circular figures.}
\begin{tabular}{llrrrrr}
\hline 
Figure  & System  & $\text{CNT}_{1}$  & $\text{CNT}_{2}$  & $\text{CNT}_{2}^{2}$  & $\text{CNT}_{3}$  & $\text{CNTF}$\tabularnewline
\hline 
Impossible  & Option 1  & $1$  & $1$  & $1$  & $1$  & $1$\tabularnewline
 & Option 2  & $1$  & $2$  & $2$  & $1$  & $1$\tabularnewline
\hline 
Realizable  & Option 1  & $0$  & $0$  & $0$  & $0$  & $0$\tabularnewline
 & Option 2  & $0$  & $0$  & $0$  & $0$  & $0$\tabularnewline
\hline 
\end{tabular}
\end{table}

\begin{table}[h]
\centering{}\caption{\label{tab:Contextuality-measures-of-4}Contextuality measures of
systems representing the Ascending and Descending staircase.}
\begin{tabular}{lrrrrr}
\hline 
Representation  & $\text{CNT}_{1}$  & $\text{CNT}_{2}$  & $\text{CNT}_{2}^{2}$  & $\text{CNT}_{3}$  & $\text{CNTF}$\tabularnewline
\hline 
Option 1  & $\nicefrac{6}{3}$  & $\nicefrac{4}{3}$  & $\nicefrac{4}{3}$  & $\nicefrac{2}{3}$  & $1$\tabularnewline
Option 2  & $1$  & $\nicefrac{9}{2}$  & $2$  & $1$  & $1$\tabularnewline
\hline 
\end{tabular}
\end{table}

\FloatBarrier 

\end{document}